\documentclass[aps,prb,twocolumn,notitlepage,groupedaddress]{revtex4-1}

\usepackage{graphicx}
\begin{document}

%\preprint{}

\title{Methods of electron transport in ab initio theory
       of spin stiffness}


\author{I. Turek}
\email[]{turek@ipm.cz}
%\thanks{}
\affiliation{Institute of Physics of Materials,
Czech Academy of Sciences,
\v{Z}i\v{z}kova 22, CZ-616 62 Brno, Czech Republic}

\author{J. Kudrnovsk\'y}
\email[]{kudrnov@fzu.cz}
\affiliation{Institute of Physics, 
Czech Academy of Sciences,
Na Slovance 2, CZ-182 21 Praha 8, Czech Republic}

\author{V. Drchal}
\email[]{drchal@fzu.cz}
\affiliation{Institute of Physics, 
Czech Academy of Sciences,
Na Slovance 2, CZ-182 21 Praha 8, Czech Republic}


\date{\today}

\begin{abstract}

We present an \emph{ab initio} theory of the spin-wave stiffness
tensor for ordered and disordered itinerant ferromagnets with pair
exchange interactions derived from a method of infinitesimal spin
rotations.
The resulting formula bears an explicit form of a linear-response
coefficient which involves one-particle Green's functions
and effective velocity operators encountered in a recent theory
of electron transport.
Application of this approach to ideal metal crystals yields
more reliable values of the spin stiffness than traditional
ill-convergent real-space lattice summations.
The formalism can also be combined with the coherent potential
approximation for an effective-medium treatment of random alloys,
which leads naturally to an inclusion of disorder-induced vertex
corrections to the spin stiffness.
The calculated concentration dependence of the spin-wave stiffness
of random fcc Ni-Fe alloys can be ascribed to a variation of the
reciprocal value of alloy magnetization.
Calculations for random iron-rich bcc Fe-Al alloys reveal that their
spin-wave stiffness is strongly reduced owing to the atomic ordering;
this effect takes place due to weakly coupled local magnetic moments
of Fe atoms surrounded by a reduced number of Fe nearest neighbors.

\end{abstract}

%\pacs{75.10.Hk, 75.30.Ds, 75.50.Pp}
%\keywords{}

\maketitle

%\subsection{}
%\subsubsection{}

\section{Introduction\label{s_intr}}

The spin stiffness, more specifically also referred to as exchange
or spin-wave stiffness, belongs undoubtedly to the most important
properties of itinerant ferromagnets.
Its value controls, e.g., the temperature dependence of magnetization
at low temperatures, the magnon dispersion law for long wavelengths,
or the width of magnetic domain walls.
Its reliable experimental or theoretical determination is thus
relevant for the whole class of ferromagnetic materials, ranging from
pure transition metals \cite{r_2016_mek} to dilute magnetic
semiconductors \cite{r_2013_nnt}.

Existing methods of \emph{ab initio} calculations of magnon spectra
and the spin stiffness, as a rule based on the density-functional
theory, include a random-phase approximation \cite{r_1981_cwl} and
techniques dealing with noncollinear magnetic structures, namely,
selfconsistent total-energy calculations of spin spirals
\cite{r_1998_lms, r_1997_rj, r_1998_hep, r_1999_bnw} and the method
of infinitesimal spin rotations \cite{r_1984_lkg, r_1987_lka}.
The latter approach leads to an effective classical
Heisenberg Hamiltonian with isotropic pair exchange interactions
between the local magnetic moments; the spin stiffness can then be
expressed as a simple real-space lattice sum.
However, the asymptotic behavior of the exchange interactions for
long interatomic distances was shown to be of the
Ruderman-Kittel-Kasuya-Yosida (RKKY) form \cite{r_2001_pkt}, which
leads to an ill-convergent behavior of the real-space sum as a
function of the cutoff distance.
A numerical technique to circumvent this problem was suggested,
which is based on an artificial damping of the pair interactions
and an extrapolation to zero damping \cite{r_2001_pkt}.
This practical solution seems to be sufficient in certain cases
\cite{r_2010_bkd, r_2019_sme}; nevertheless, a more fundamental
approach to overcome this obstacle would be highly desirable.

Theory and calculations of the spin stiffness for random alloys
(substitutionally disordered systems on nonrandom crystalline
lattices) face other difficulties.
The most direct way to handle the randomness seems to be a
generalization of the total-energy spin-spiral calculations using,
e.g., the Korringa-Kohn-Rostoker (KKR) multiple-scattering theory
and the coherent potential approximation (CPA) \cite{r_2011_mfe}.
This approach neglects effects of fluctuating local environments; 
the same neglect is inherent to the method of infinitesimal spin
rotations applied to random alloys in the CPA \cite{r_2006_tkd,
r_2008_kdb}.
The local environment effect can be treated in supercell
calculations; a study performed for an equiconcentration fcc Ni-Fe
alloy proved that the CPA-averaged exchange interactions agree
reasonably well with averages from a 16-atom supercell
\cite{r_2005_rko}.
Another problem related to this topic is the correct CPA average
of the pair exchange interactions.
The latter involve a product of two one-particle resolvents
(Green's functions), so that the proper configuration average
should consist of a coherent contribution and the vertex
corrections \cite{r_1969_bv}. 
However, the vertex corrections are typically ignored in existing
studies, which is sometimes loosely justified by the so-called
vertex-cancellation theorem \cite{r_1996_bkd} relevant for
interlayer exchange coupling of two ferromagnetic layers separated
by a thick nonmagnetic spacer layer. 
Hence, the role of the vertex corrections in bulk alloy systems
deserves more detailed investigation, at least on the CPA level.

Besides the mentioned problems in determination of the exchange
interactions of random alloys, evaluation of magnon spectra
of these systems is considered as another challenge for the
solid-state theory \cite{r_2018_btm}.
The formulation of effective-medium approaches is rather
sophisticated and it leads to nontrivial numerical implementation
\cite{r_2002_bb, r_2016_bsb}.
Various brute-force simulations using large supercells have thus been
used as alternatives which may be efficient and reliable in a number
of cases, including both the spin-wave spectra and values of the
spin-wave stiffness \cite{r_2016_bsb, r_2008_shn, r_2007_gb,
r_2016_tkd, r_2018_btm}.
It was found that the spin stiffness of random diluted ferromagnets
is reduced due to the sites without local magnetic moments as
compared to the value obtained for a clean crystal with nonrandom
concentration-weighted exchange interactions of the alloy.
This reduction is weak for small concentrations of magnetic
vacancies \cite{r_2016_bsb}, but it becomes appreciable in systems
with strong dilution, such as, e.g., Mn-doped GaAs \cite{r_2007_gb,
r_2016_tkd}.

The spin (exchange) stiffness represents one of the micromagnetic
parameters of a ferromagnet which describe the energetics and
dynamics in cases with magnetization direction slowly varying
in space (i.e., the magnetization variations are featured by
length scales substantially exceeding the nearest-neighbor
interatomic distance).
Another micromagnetic parameter, sometimes called spiralization
\cite{r_2014_fbm_c}, is due to relativistic effects, especially
the spin-orbit coupling and the closely related
Dzyaloshinskii-Moriya interaction.
An \emph{ab initio} relativistic theory of spin stiffness and
spiralization for ordered and disordered systems has recently been
developed within the KKR-CPA technique \cite{r_2019_mpe}
as an extension of the method of infinitesimal spin rotations
\cite{r_1984_lkg, r_1987_lka}.
For clean crystals, the spiralization was also formulated in terms
of the Berry phase of $\mathbf{k}$-vector dependent Bloch
eigenstates of the effective one-electron Hamiltonian
\cite{r_2014_fbm_c}, i.e., using a concept encountered
in the theory of electron transport properties such as the anomalous
Hall conductivity \cite{r_2010_nso}.
A natural question arises thus in this context, namely, whether the
spin stiffness can also be expressed
as a linear-response coefficient similar to the conductivity
and evaluated by means of techniques employed for electron transport,
with applicability to random alloys as well.

From the materials point of view, existing applications of the
current \emph{ab initio} techniques for the spin stiffness were
focused on pure ferromagnetic $3d$ transition metals (Fe, Co, Ni)
\cite{r_1997_rj, r_1999_bnw, r_2001_pkt, r_2016_mek, r_2020_sme},
selected stoichiometric ordered compounds \cite{r_2014_kdb,
r_2010_bkd}, random binary and ternary transition-metal alloys
\cite{r_2011_mfe, r_2019_mpe, r_2019_sme, r_2020_sme}, and dilute
magnetic semiconductors (Mn-doped GaAs) \cite{r_2007_gb, r_2016_tkd}.
Less attention has been devoted so far to random alloys of transition
metals with $p$ elements, such as, e.g., bcc Fe-M substitutional
solid solutions, where M = Be, Al, Si, and Ga. 
Some of these iron-rich alloys exhibit pronounced magnetoelastic
properties (tetragonal magnetostriction) which motivated a number
of experimental studies \cite{r_2005_zgl, r_2006_zcw}.
Full assessment of the microscopic origin of this behavior requires
fair knowledge of the phonon and magnon spectra.
As a rule, the measured magnon spectra of the mentioned bcc Fe-M
alloys point to magnon softening due to M alloying \cite{r_2007_zml}.
However, a recent \emph{ab initio} study of the spin-wave stiffness
of random Fe-Al alloys indicates an opposite concentration trend
\cite{r_2016_bsb}; this qualitative discrepancy deserves thus closer
examination. 

The main aim of this study is to present an alternative
formalism for the calculation of the spin stiffness in nonrandom and
random ferromagnetic systems which employs current techniques of
electron transport theory. 
The developed scheme is then used to address some of the
above-mentioned methodological and physical problems.
The paper is organized as follows:
the theoretical formalism is introduced in Section~\ref{s_thfo},
the numerical details are listed in Section~\ref{s_nuim},
and the results are discussed in Section~\ref{s_redi},
including those for pure ferromagnetic $3d$ transition metals
(Section~\ref{ss_ptm}), 
for random fcc Ni-Fe alloys (Section~\ref{ss_nife}),
and for random bcc Fe-Al alloys (Section~\ref{ss_feal}).
Concluding remarks are presented in Section~\ref{s_conc}.

%\clearpage

\section{Theoretical formalism\label{s_thfo}}

The starting point of our approach to the spin stiffness is the
classical Heisenberg Hamiltonian
\begin{equation}
\mathcal{E}( \{ \mathbf{e}_\mathbf{R} \} ) = 
- \sum_{\mathbf{R}\mathbf{R}'} J_{\mathbf{R}\mathbf{R}'} 
\mathbf{e}_\mathbf{R} \cdot \mathbf{e}_{\mathbf{R}'} ,
\label{eq_chm}
\end{equation}
where the indices $\mathbf{R}$ and $\mathbf{R}'$ label the lattice
sites, the unit vectors $\mathbf{e}_\mathbf{R}$ denote directions
of local moments attached to respective lattice sites, and the
quantities $J_{\mathbf{R}\mathbf{R}'}$ are pair exchange interactions
(satisfying $J_{\mathbf{R}\mathbf{R}} = 0$ and
$J_{\mathbf{R}\mathbf{R}'} = J_{\mathbf{R}'\mathbf{R}}$).
This Hamiltonian is appropriate for ferromagnetic systems with
neglect of relativistic effects and for local-moment directions
deviating only slightly from the ground-state magnetization
direction.
The latter direction is assumed along the $z$ axis in the following.
The spin stiffness is related naturally to energies of spin spirals
which are parametrized by a reciprocal space vector $\mathbf{q}$
and a cone angle $\theta$.
The spin structure of the spin spiral is then defined explicitly as
\begin{equation}
\mathbf{e}_\mathbf{R} = \left( 
\sin \theta \cos ( \mathbf{q} \cdot \mathbf{R} ) ,
\sin \theta \sin ( \mathbf{q} \cdot \mathbf{R} ) ,
\cos \theta \right) ,
\label{eq_ssdef}
\end{equation}
which yields
\begin{equation}
\mathbf{e}_\mathbf{R} \cdot \mathbf{e}_{\mathbf{R}'}
 = \cos^2 \theta + \sin^2 \theta 
\cos [ \mathbf{q} \cdot ( \mathbf{R} - \mathbf{R}' ) ] .
\label{eq_spq}
\end{equation}
The energy of the spin spiral with respect to that of the
ferromagnetic ground state is then equal to
\begin{equation}
\delta E ( \theta, \mathbf{q} ) = \sin^2 \theta 
\sum_{\mathbf{R}\mathbf{R}'} J_{\mathbf{R}\mathbf{R}'} 
\left\{ 1 - \cos [ \mathbf{q} \cdot 
( \mathbf{R} - \mathbf{R}' ) ] \right\} .
\label{eq_dess}
\end{equation}
This expression in the limit $| \mathbf{q} | = q \to 0$ can be
used for a definition of the exchange stiffness relevant, e.g.,
for energetics of domain walls.

For the spin-wave stiffness related to the magnon spectra and
considered in the rest of this paper, one has to include the
total spin magnetic moment $M$ of the solid and the quantization
of $z$ component of the total spin operator \cite{r_1997_rj}.
This leads to a condition for the small cone angle $\theta$
given by
\begin{equation}
2 \mu_\mathrm{B} = \delta m_z = M ( 1 - \cos \theta )
\approx \frac{1}{2} M \theta^2 ,
\label{eq_deltamz}
\end{equation}
where the gyromagnetic ratio $g=2$ for electrons is assumed,
$\mu_\mathrm{B}$ denotes the Bohr magneton, and $\delta m_z$ is
the change of $z$ component of the total magnetic moment. 
The last condition together with Eq.~(\ref{eq_dess}) yields the
magnon energy
\begin{eqnarray}
E_\mathrm{mag} (\mathbf{q}) & = & \frac{4\mu_\mathrm{B}}{M} 
\sum_{\mathbf{R}\mathbf{R}'} J_{\mathbf{R}\mathbf{R}'} 
\left\{ 1 - \cos [ \mathbf{q} \cdot 
( \mathbf{R} - \mathbf{R}' ) ] \right\} 
\nonumber\\
 & \approx & \sum_{\mu\nu}
D_{\mu\nu} q_\mu q_\nu ,
\label{eq_emag}
\end{eqnarray}
where the approximate relation is valid for small $\mathbf{q}$
vectors with Cartesian components $q_\mu$ ($\mu \in \{ x , y , z \}$)
and where $D_{\mu\nu}$ denotes the spin-wave stiffness tensor.
This tensor is explicitly given by
\begin{equation}
D_{\mu\nu} = \frac{2\mu_\mathrm{B}}{M} 
 \sum_{\mathbf{R}\mathbf{R}'} J_{\mathbf{R}\mathbf{R}'} 
( X^\mu_\mathbf{R} - X^\mu_{\mathbf{R}'} )
( X^\nu_\mathbf{R} - X^\nu_{\mathbf{R}'} ) ,
\label{eq_dmndef}
\end{equation}
where the symbol $X^\mu_\mathbf{R}$ denotes the $\mu$ component
of the lattice site vector $\mathbf{R}$.

According to the well-known formalism based on infinitesimal spin
rotations and the magnetic force theorem \cite{r_1987_lka}, the pair
exchange interactions $J_{\mathbf{R}\mathbf{R}'}$ can be expressed in
terms of the electronic structure of the ferromagnetic ground state
as
\begin{eqnarray}
J_{\mathbf{R}\mathbf{R}'} & = &
\frac{i}{8\pi} \int_C \mathrm{tr}_L
\left\{ \Delta_\mathbf{R}(z) 
g^\uparrow_{\mathbf{R}\mathbf{R}'}(z)
\Delta_{\mathbf{R}'}(z) 
g^\downarrow_{\mathbf{R}'\mathbf{R}}(z)
\right\} dz ,
\nonumber\\
 & & \Delta_\mathbf{R}(z) =
P^\uparrow_\mathbf{R}(z) - P^\downarrow_\mathbf{R}(z) .
\label{eq_jrrp}
\end{eqnarray}
In this relation, the trace ($\mathrm{tr}_L$) is taken over the
composed orbital index $L = (\ell , m)$, the argument $z$ denotes a
complex energy variable, the complex integration contour $C$ is
oriented counterclockwise, with starting and ending point at the
Fermi energy $E_\mathrm{F}$ and containing the whole occupied valence
spectrum.
The quantities $g^s_{\mathbf{R}\mathbf{R}'}(z)$ abbreviate blocks
of matrix elements $g^s_{\mathbf{R}L,\mathbf{R}'L'} (z)$ of
one-electron Green's functions $g^s(z)$ in the spin channel $s$,
where $s \in \{ \uparrow , \downarrow \}$ is the spin index.
The complex integration in Eq.~(\ref{eq_jrrp}) is equivalent to the
standard real-energy integration \cite{r_1987_lka, r_2006_tkd} owing
to analyticity of the integrated function.
In this work, we employ the linear muffin-tin orbital (LMTO) method
\cite{r_1975_oka, r_1983_gja, r_1997_tdk},
in which the $g^s(z)$ refers to the auxiliary Green's function
defined by
\begin{equation}
g^s(z) = [ P^s(z) - S ]^{-1} ,
\label{eq_gsz}
\end{equation}
where $P^s(z)$ denotes the site-diagonal matrix of potential
functions and $S$ is the LMTO structure-constant matrix.
The site-diagonal blocks $P^s_\mathbf{R}(z)$ of the matrices $P^s(z)$
define the energy-dependent local exchange splittings
$\Delta_\mathbf{R}(z)$ entering the expression for 
$J_{\mathbf{R}\mathbf{R}'}$ (\ref{eq_jrrp}); the blocks 
$\Delta_\mathbf{R}(z)$ form a site-diagonal matrix $\Delta(z) =
P^\uparrow(z) - P^\downarrow(z)$ to be used in the following. 
Let us note that the LMTO formalism employed here can be replaced by
the KKR formalism, which leads to a replacement of the auxiliary
Green's function $g^s(z)$ in the last two equations by the
scattering path operator \cite{r_1990_pw, r_2005_zhs}.

The formulation of a compact expression for the spin-wave stiffness
tensor $D_{\mu\nu}$ rests on the definition of coordinate operators
$X^\mu$ ($\mu \in \{ x , y , z \}$), represented by matrices diagonal
in the site ($\mathbf{R}$) and orbital ($L$) indices, given explicitly
by
\begin{equation}
X^\mu_{\mathbf{R}L,\mathbf{R}'L'} = 
\delta_{\mathbf{R}\mathbf{R}'} \delta_{LL'} X^\mu_\mathbf{R} .
\label{eq_xmu}
\end{equation}
These coordinate operators were introduced in an \emph{ab initio}
theory of electron transport \cite{r_2002_tkd}.
In the present context, one can use them in relations of the type
\begin{equation}
g^s_{\mathbf{R}L,\mathbf{R}'L'} (z)
\left( X^\mu_\mathbf{R} - X^\mu_{\mathbf{R}'} \right) = 
\left[ X^\mu , g^s(z) \right]_{\mathbf{R}L,\mathbf{R}'L'} ,
\label{eq_mecomm}
\end{equation}
where $[A,B] = AB - BA$ is a commutator.
The last relation together with Eqs.~(\ref{eq_dmndef}) and
(\ref{eq_jrrp}) lead to the tensor $D_{\mu\nu}$ in the form of
a contour integral
\begin{equation}
D_{\mu\nu} = \frac{\mu_\mathrm{B}}{2M} \, \frac{1}{2\pi i}
\int_C f_{\mu\nu}(z) dz ,
\label{eq_dmn}
\end{equation}
with the integrated function $f_{\mu\nu}(z)$ given by
\begin{equation}
f_{\mu\nu} = \mathrm{Tr} \{ \Delta [ X^\mu , g^\uparrow ]
\Delta [ X^\nu , g^\downarrow ] \} ,
\label{eq_fmnst}
\end{equation}
where all energy arguments (equal $z$) have been suppressed for
brevity and where the trace (Tr) extends over all $\mathbf{R}L$
indices of the whole system.
The commutators in the last relation can be rewritten as
\begin{eqnarray}
[ X^\mu , g^s ] & = & i g^s v_\mu g^s ,
\nonumber\\
v_\mu & = & -i [ X^\mu , S ] ,
\label{eq_commxg}
\end{eqnarray}
where we introduced the effective velocity operators $v_\mu$ that
enter the LMTO transport theory as well \cite{r_2002_tkd, r_2012_tkd,
r_2014_tkd}.
The relation (\ref{eq_commxg}) follows from Eq.~(\ref{eq_gsz}) and
from the site-diagonal nature of the potential functions $P^s(z)$,
which implies $[ X^\mu , P^s(z) ] = 0$.
The substitution of Eq.~(\ref{eq_commxg}) into Eq.~(\ref{eq_fmnst})
and the use of cyclic property of trace together with an identity
\begin{equation}
g^\uparrow \Delta g^\downarrow =
g^\downarrow \Delta g^\uparrow =
g^\downarrow - g^\uparrow 
\label{eq_gdg}
\end{equation}
yield the final expression for the function $f_{\mu\nu}(z)$, namely,
\begin{equation}
f_{\mu\nu} = - \mathrm{Tr} \{ 
 v_\mu ( g^\uparrow - g^\downarrow ) 
 v_\nu ( g^\uparrow - g^\downarrow ) \} ,
\label{eq_fmnfin}
\end{equation}
where energy arguments $z$ are omitted.
This is the central result of this section.

The final expression for the spin stiffness tensor, 
see Eq.~(\ref{eq_dmn}) and Eq.~(\ref{eq_fmnfin}), has a form of a
genuine linear-response coefficient suitable for direct numerical
evaluation. 
This calculation requires merely the selfconsistent electronic
structure of the ferromagnetic ground state as an input for the
relevant integrations over the Brillouin zone (BZ) and over the
complex energy path (Section~\ref{s_nuim}).
This straightforward procedure should be contrasted with most of
existing approaches which require (in addition to the selfconsistent
ferromagnetic ground state) another intermediate step which is
numerically quite demanding or delicate. 
This refers to the method based on the ill-convergent real-space
lattice summation \cite{r_2001_pkt} (where the real-space pair
exchange interactions have to be obtained first, followed by an
extrapolation with respect to the artificial damping parameter),
to the technique employing the spin-spiral calculations
\cite{r_1997_rj} (where total energies of the spin spirals for finite
$\mathbf{q}$ vectors and cone angles $\theta$ have to be obtained
first, followed by numerical derivatives), as well as to the recent
KKR approach with reciprocal-space integration \cite{r_2019_mpe}
(where numerically demanding derivatives of the scattering-path
operator with respect to the $\mathbf{k}$-vector components have to
be evaluated).

The derived alternative formula for the spin-wave stiffness tensor
deserves also further comments.
First, in contrast to local exchange splittings entering the 
previous expressions for this tensor \cite{r_1984_lkg, r_2019_mpe},
the present result contains the nonlocal, spin-independent velocity
operators $v_\mu$ while all effects of the exchange splitting are
contained in the difference $g^\uparrow(z) - g^\downarrow(z)$ of the
spin-resolved Green's functions.
Second, the form of $f_{\mu\nu}(z)$ (\ref{eq_fmnfin}) resembles
strongly the Kubo-Greenwood formula for the conductivity tensor
$\sigma_{\mu\nu}$ in the LMTO method \cite{r_2002_tkd}.
The latter is obtained by replacing $g^\uparrow(z)$ and
$g^\downarrow(z)$ in Eq.~(\ref{eq_fmnfin}) by $g(E_\mathrm{F} + i0)$
and $g(E_\mathrm{F} - i0)$, respectively, i.e., by the retarded and
advanced Green's functions at the Fermi energy.
Third, this analogy with the theory of electron transport enables one
to apply the same techniques of configuration averaging in the CPA
also for the spin-wave stiffness tensor of random alloys.
In particular, the quantity $M$ in Eq.~(\ref{eq_dmn}) has to be
replaced by the average alloy magnetization and the CPA average
of the Green's function \cite{r_1997_tdk, r_1990_kd}
\begin{equation}
\bar{g}^s(z) = [ \mathcal{P}^s(z) - S ]^{-1} ,
\label{eq_gav}
\end{equation}
where $\mathcal{P}^s(z)$ denotes the site-diagonal matrix of
coherent potential functions, is used for the average of
Eq.~(\ref{eq_fmnfin}).
This leads to the result
\begin{equation}
\bar{f}_{\mu\nu}(z) = \bar{f}^\mathrm{coh}_{\mu\nu}(z)
                    + \bar{f}^\mathrm{VC}_{\mu\nu}(z) ,
\label{eq_fmnav}
\end{equation}
where the first term defines the coherent (coh) part, given
explicitly by
\begin{equation}
\bar{f}^\mathrm{coh}_{\mu\nu} = - \mathrm{Tr} \{ 
 v_\mu ( \bar{g}^\uparrow - \bar{g}^\downarrow ) 
 v_\nu ( \bar{g}^\uparrow - \bar{g}^\downarrow ) \} , 
\label{eq_fmncoh}
\end{equation}
while the second term in Eq.~(\ref{eq_fmnav}) is the corresponding
incoherent part (vertex corrections, VC).
Note that this decomposition follows the original approach by
Velick\'y \cite{r_1969_bv} owing to the nonrandom effective
velocities $v_\mu$; the vertex corrections in the present
LMTO-CPA formalism have been evaluated according to the Appendix
to Ref.~\onlinecite{r_2006_ctk}.
As a direct consequence of the decomposition (\ref{eq_fmnav}), the
spin-wave stiffness tensor of random alloys can also be written as
a sum of its coherent and incoherent parts, $D_{\mu\nu} = 
D^\mathrm{coh}_{\mu\nu} + D^\mathrm{VC}_{\mu\nu}$, which represents
the complete CPA average.
Fourth, the proposed approach is not limited to the LMTO technique,
but it is transferable to other electronic structure methods, such
as the KKR technique.
The effective velocities $v_\mu$ enter the formalism via the commutator
relation, Eq.~(\ref{eq_commxg}), which involves the diagonal coordinate
operators $X^\mu$, Eq.~(\ref{eq_xmu}).
The transferability rests on the very simple form of these coordinate
operators which reflects the basic starting point of the method of
infinitesimal spin rotations \cite{r_1984_lkg, r_1987_lka}, in which
the rotations refer to the whole local magnetic moments at the
respective lattice sites.

Finally, the original (canonical) LMTO formalism used in this
section can be replaced by its tight-binding (TB) version, in which
the potential functions $P^s(z)$, the structure-constant matrix $S$,
and the Green's functions $g^s(z)$ are replaced by their screened
counterparts \cite{r_1984_aj, r_1986_apj, r_1997_tdk}.
The TB-LMTO technique is advantageous for numerical implementation.
It can be proved that the function $f_{\mu\nu}(z)$ (\ref{eq_fmnfin})
and the tensor $D_{\mu\nu}$ (\ref{eq_dmn}) are invariant with
respect to the TB-LMTO screening transformation.
This invariance holds also within the CPA, so that both
$D^\mathrm{coh}_{\mu\nu}$ and $D^\mathrm{VC}_{\mu\nu}$ are invariant
quantities as well.
The proof of invariance is omitted here for its similarity to that
done in the case of transport properties \cite{r_2014_tkd}.

\section{Numerical implementation\label{s_nuim}}

The developed theory was implemented numerically in a way resembling
that employed recently for the so-called Fermi sea contribution
to the conductivity tensor in the relativistic TB-LMTO-CPA theory
\cite{r_2014_tkd}.
The selfconsistent electronic structure was obtained within the local
spin-density approximation (LSDA) and the atomic sphere approximation
using the scalar-relativistic $spd$ basis \cite{r_1997_tdk}.
The contour integration in Eq.~(\ref{eq_dmn}) was performed along
a circular path $C$ of a diameter 1.5~Ry; the numerical integration
was done with 20 to 40 complex nodes distributed on the lower half
of $C$.
The BZ averages were evaluated by using $\sim 10^8$ $\mathbf{k}$
vectors for a uniform sampling of one half of the full BZ.

For the sake of a comparison of the $D_{\mu\nu}$ calculated from
Eq.~(\ref{eq_dmn}) with results of traditional approaches, we have
also applied a procedure based on TB-LMTO-CPA total-energy
calculations for spin spirals (\ref{eq_ssdef}), implemented according
to the KKR-CPA technique \cite{r_2011_mfe}.
The calculations were performed for cubic systems with planar spirals
($\theta = \pi/2$) and $\mathbf{q}$ vectors along the $z$ axis,
$\mathbf{q} = (0,0,q)$.
The spin-wave stiffness, denoted as $D_\mathrm{sp}$ in the following,
was then obtained from a numerical derivative of the total energy as
a function of $q$ for $q \to 0$ \cite{r_1997_rj}.

%\clearpage

\section{Results and discussion\label{s_redi}}

\subsection{Pure transition metals\label{ss_ptm}}

\begin{figure}[htb]
\begin{center}
\includegraphics[width=0.85\columnwidth]{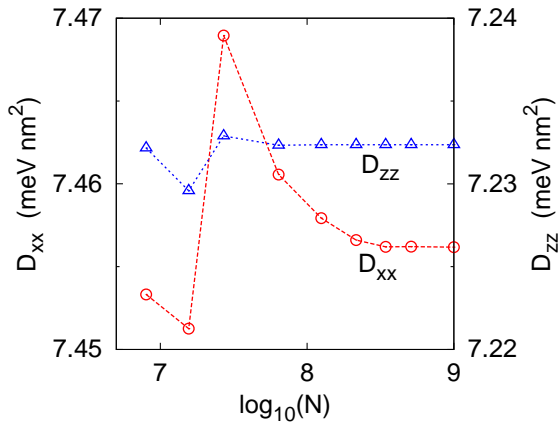}
\end{center}
\caption{
The calculated spin-wave stiffnesses $D_{xx} = D_{yy}$
(circles, left scale) and $D_{zz}$ (triangles, right scale)
for hcp Co as functions of the number $N$ of $\mathbf{k}$
vectors sampling one half of the first Brillouin zone.
\label{f_co}}
\end{figure}

The numerical aspects of the developed formalism have been first
examined for pure ferromagnetic $3d$ transition metals: bcc Fe, 
hcp Co, and fcc Ni.
Figure~\ref{f_co} shows the convergence behavior of the spin-wave
stiffness tensor for hcp Co.
One can see that with the increasing number $N$ of $\mathbf{k}$
vectors sampling the hcp BZ, both nonzero elements of $D_{\mu\nu}$,
namely $D_{xx} = D_{yy}$ and $D_{zz}$, exhibit a fairly rapid
convergence to their limiting values.
A similar fast convergence has been observed for cubic iron and
nickel (not shown here).
This convergence property is substantially better than that of the
real-space lattice summations (involving the pair exchange
interactions) as functions of the cutoff distance $d_\mathrm{max}$,
see Fig.~2 of Ref.~\onlinecite{r_2001_pkt}.
The reason lies in the typical modest values of $d_\mathrm{max} <
 10a$ used in the lattice summations, where $a$ denotes the lattice
parameter.
The employed numbers $N > 10^6$ for the BZ sampling (Fig.~\ref{f_co})
are equivalent to big but finite crystals (with periodic boundary
conditions) with edge lengths $L \approx N^{1/3} a > 100 a$,
exceeding thus the cutoff distances $d_\mathrm{max}$ by at least one
order of magnitude.

Let us note that the convergence problems of the real-space lattice
summations are caused by the RKKY-like asymptotic behavior of the pair
exchange interactions, leading to their very slow decay for long
intersite distances.
The new formalism removes the real-space pair interactions completely
by using the BZ integrations. 
Its efficiency is closely related to the lattice Fourier transformation
of the TB-LMTO structure constant matrix $S$, which is the only
non-site-diagonal matrix entering the evaluated expression.
The spatial range of the matrix $S$ is extremely short, so that
typically a cutoff to the second or third shell of nearest neighbors is
fully sufficient for close-packed lattices such as fcc, bcc, or hcp
\cite{r_1984_aj, r_1986_apj}.
This property allows one to perform the Fourier transformation very
fast and, consequently, to increase the number of the sampling
$\mathbf{k}$ vectors substantially.

\begin{table}[htb]
\caption{The calculated and experimental values of the spin-wave
stiffness $D$ (in meV nm$^2$) for $3d$ transition-metal ferromagnets.
The values obtained in this work are completed by values calculated
previously from spin spirals \cite{r_1997_rj} and from real-space
lattice summations \cite{r_2001_pkt}.
\label{t_dpure}}
\begin{ruledtabular}
\begin{tabular}{lcccc}
      & Calc.$^\mathrm{a}$ & Calc.$^\mathrm{b}$ & 
  Calc.$^\mathrm{c}$ & Exp.  \\
\hline
Fe &  2.73  &  2.47  &  2.50  &  3.3$^\mathrm{d}$  \\
Co &  7.38  &  5.02  &  6.63  &  5.8$^\mathrm{e}$  \\
Ni &  8.01  &  7.39  &  7.56  &  5.3$^\mathrm{f}$  \\
\end{tabular}
\end{ruledtabular}
$^\mathrm{a}$This work. \
$^\mathrm{b}$Reference \onlinecite{r_1997_rj}. \
$^\mathrm{c}$Reference \onlinecite{r_2001_pkt}. \
$^\mathrm{d}$Reference \onlinecite{r_1964_hhl}. \
$^\mathrm{e}$Reference \onlinecite{r_1982_rp_b}. \
$^\mathrm{f}$Reference \onlinecite{r_1983_in}. 
\end{table}

The converged values of the spin-wave stiffness $D$ for all three
metals are summarized in Table~\ref{t_dpure} together with values
obtained from the spin-spiral calculations \cite{r_1997_rj} and the
real-space lattice summations \cite{r_2001_pkt} as well as from
experiments.
The value of $D$ for hcp Co in the table refers to the isotropic
part of $D_{\mu\nu}$, i.e., $D = (2D_{xx} + D_{zz})/3$, whereas the
values from both previous calculations \cite{r_1997_rj, r_2001_pkt}
refer to fcc Co; the experimental value \cite{r_1982_rp_b} was
obtained for the hcp phase.
One can observe that all theoretical values reproduce roughly the
measured data with the biggest discrepancy encountered for nickel;
for the cubic metals (Fe and Ni), similar values of
$D$ have recently been obtained by a thorough analysis of results of
the KKR multiple-scattering theory \cite{r_2020_sme}.
The overestimation of $D$ for Ni by the LSDA calculations can be
explained by the well-known overestimation of the exchange
splitting which comes out about two times bigger than from
photoemission experiments \cite{r_1980_ehk}.
A systematic way to achieve better agreement between the theory and
the experiment should include effects of electron-electron
correlations beyond the LSDA \cite{r_2000_kl}; this task has to be
left for future studies.

The calculated tensor $D_{\mu\nu}$ for hcp Co exhibits a small
anisotropy featured by $D_{xx} > D_{zz}$ (Fig.~\ref{f_co}).
However, a recent calculation based on the pair exchange interactions
up to six nearest neighbors \cite{r_2016_mek} yields an opposite
anisotropy ($D_{xx} < D_{zz}$).
This fact documents the importance of well converged values for
reliable resolution of subtle details of the spin-wave stiffness
tensor.
Similarly, the values of $D$ for bcc Fe and fcc Ni obtained in this
work (Table~\ref{t_dpure}) differ from those based on the real-space
lattice summation \cite{r_2001_pkt}; the relative difference (below
10\% for both metals) points to an uncertainty inherent to the
employed regularization procedure of the ill-convergent lattice sums
\cite{r_2001_pkt}.
Let us note that pure metals and ordered clean crystals represent
the most difficult cases for the lattice-sum approach, whereas random
alloys can be treated by this technique with a higher accuracy and
efficiency due to the disorder-induced exponential damping of
the exchange interactions for large interatomic distances
\cite{r_2019_sme}.

%\clearpage

\subsection{Random fcc Ni-Fe alloys\label{ss_nife}}

\begin{figure}[htb]
\begin{center}
\includegraphics[width=0.90\columnwidth]{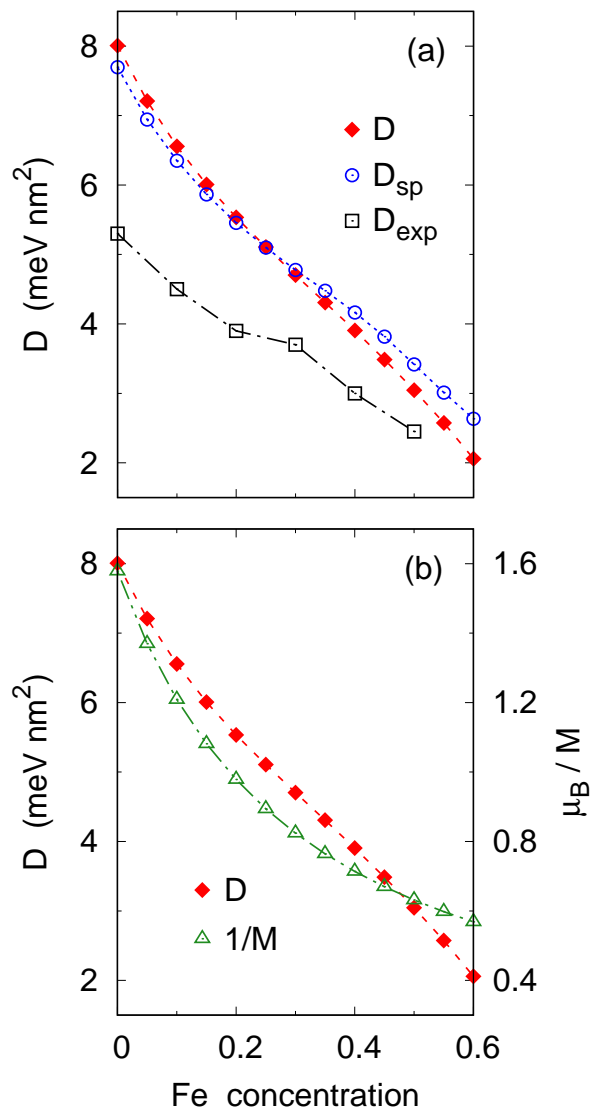}
\end{center}
\caption{
Concentration dependences of the spin-wave stiffness and
the alloy magnetization in random fcc Ni-Fe alloy:
(a) the stiffness from Eq.~(\ref{eq_dmn}) ($D$, solid diamonds),
from spin-spiral calculations ($D_\mathrm{sp}$, open circles), and
from experiment \cite{r_1983_in} ($D_\mathrm{exp}$, open boxes);
(b) the stiffness from Eq.~(\ref{eq_dmn}) ($D$, solid diamonds,
left scale) and the reciprocal value of magnetization per atom
($1/M$, open triangles, right scale).
\label{f_nife}}
\end{figure}

For the random fcc Ni-Fe alloys, we considered a concentration
range up to 60\% Fe; the fcc lattice parameter for a given alloy
composition was set according to the Vegard's law using the
experimental atomic (Wigner-Seitz) radii of the pure constituents
in their equilibrium structures, $s_\mathrm{Ni} = 2.60 \, a_0$ and
$s_\mathrm{Fe} = 2.66 \, a_0$, where $a_0$ denotes the Bohr radius.
The calculated spin-wave stiffness $D$ is displayed in 
Fig.~\ref{f_nife}(a) together with values from the spin-spiral
calculations ($D_\mathrm{sp}$) and from the experiment
\cite{r_1983_in}.
One can see that both calculated quantities, $D$ and $D_\mathrm{sp}$,
acquire mutually close values and exhibit very similar concentration
trends, giving thus confidence in both formalisms and their numerical
implementations.
Moreover, the calculated values for the permalloy composition
Ni$_{0.8}$Fe$_{0.2}$, namely $D = 5.53$ meV nm$^2$ and
$D_\mathrm{sp} = 5.45$ meV nm$^2$, compare reasonably well with
recent KKR values for a Ni$_{0.81}$Fe$_{0.19}$ alloy, which lie in
an interval $[ 5.12 , 5.63 ]$ meV nm$^2$ depending on the particular
approach employed \cite{r_2020_sme}.
The experimental stiffness shows also a similar decreasing trend
with increasing Fe concentration; however, the measured values are
appreciably smaller than the theoretical ones, especially for 
Ni-rich alloys, which originates in the discrepancy found for pure
Ni, see Section~\ref{ss_ptm}.

The theoretical results in Fig.~\ref{f_nife}(a) are in reasonable
agreement with those obtained recently from a fully relativistic
extension of the method of infinitesimal spin rotations, see Fig.~1
in Ref.~\onlinecite{r_2019_mpe}.
This fact indicates that the spin-orbit interaction has a negligible
effect on the spin stiffness in this alloy system and that its
omission in the present work cannot be responsible for the existing
discrepancy between the theory and experiment for Ni-rich alloys.
The decreasing trend of $D$ with increasing Fe content deserves a
brief comment as well, since attempts to explain similar
concentration dependences in binary transition-metal systems appeared
rather early \cite{r_1964_hhl, r_1965_lss}.
Fig.~\ref{f_nife}(b) displays the calculated value of $D$ together
with the reciprocal value of the alloy magnetization $M$, which
enters the expressions for $D$, see Eqs.~(\ref{eq_dmndef}) and
(\ref{eq_dmn}).
While the concentration trends of $D$ and $1/M$ differ slightly, the
largest part of the variation of the spin-wave stiffness throughout
the whole concentration range studied can safely be ascribed to the
variation of the alloy magnetization.
Finally, we note that the total stiffness $D$ in the fcc Ni-Fe alloys
coincides practically with its coherent part while the incoherent
part (vertex corrections) is completely negligible in the entire
concentration interval (the maximum vertex part is encountered for
60\% Fe, where it amounts to about 0.5\% of the total $D$). 

%\clearpage

\subsection{Random bcc Fe-Al alloys\label{ss_feal}}

\begin{figure}[htb]
\begin{center}
\includegraphics[width=0.85\columnwidth]{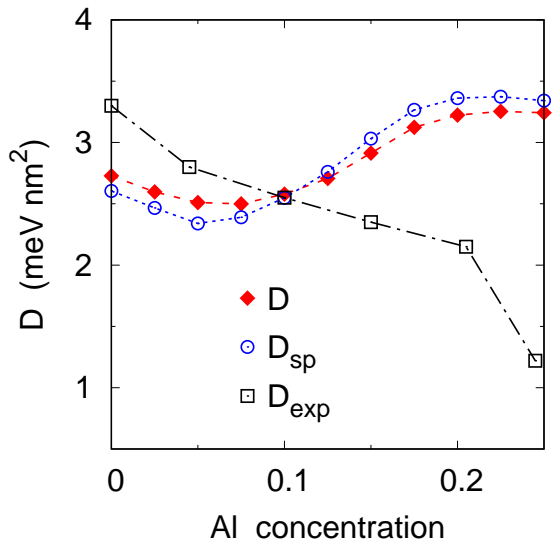}
\end{center}
\caption{
Concentration dependence of the spin-wave stiffness
in random bcc Fe-Al alloy:
the stiffness from Eq.~(\ref{eq_dmn}) ($D$, solid diamonds),
from spin-spiral calculations ($D_\mathrm{sp}$, open circles),
and from experiment \cite{r_1966_as, r_1964_hhl}
($D_\mathrm{exp}$, open boxes).
\label{f_feal_dis}}
\end{figure}

The random bcc Fe-Al alloys were studied for Al concentrations
up to 25\%; the variation of the bcc lattice parameter with
composition was set according to existing experimental data
\cite{r_1992_yvb, r_1997_kg}.
The theoretical values of the spin-wave stiffness are shown in
Fig.~\ref{f_feal_dis} simultaneously with the experimental points
\cite{r_1966_as} (the measured value for pure iron was taken
from Ref.~\onlinecite{r_1964_hhl}).
One reveals a close mutual similarity of both calculated values
($D$ and $D_\mathrm{sp}$), which however differ significantly from
the measured values.
The measured monotonic decrease of the spin-wave stiffness of bcc
Fe due to an alloying by a $p$ element M has been reported not
only for M = Al, but also for other dopants (M = Be, Ga, Si), see
Ref.~\onlinecite{r_2007_zml} and references therein.
On the theoretical side, the initial decrease of $D$ (up to 5\% Al)
is changed into an increase for higher Al contents (with a saturation
close to 25\% Al), see Fig.~\ref{f_feal_dis}.
A nonmonotonic concentration dependence of $D$ has also been
calculated by the authors of Ref.~\onlinecite{r_2016_bsb} with
a maximum stiffness around 20\% Al. 

For an explanation of the above discrepancy, one has to consider
effects of atomic ordering, pronounced in the Fe-Al alloy especially
for higher Al concentrations \cite{r_1986_tbm}, but neglected in the
calculations reported in Ref.~\onlinecite{r_2016_bsb} as well as in
our approach (Fig.~\ref{f_feal_dis}).
As suggested by several authors \cite{r_2007_zml, r_1966_as,
r_1978_ef}, the strong reduction of the spin-wave stiffness for
the alloy with 25\% Al as compared to that of pure iron should be
ascribed to the $D0_3$ or $B2$ atomic orders.
Moreover, theoretical investigation of the atomic short-range order
in bcc Fe-Al alloys predicts that the alloys with about 20\% Al
exhibit a substantial degree of the B2 short-range order when
annealed from high temperatures \cite{r_1997_slj, r_1998_srl}.
Our theoretical spin-wave stiffness for the stoichiometric Fe$_3$Al
system with the $D0_3$ structure amounts to $D = 1.74$ meV nm$^2$,
which represents a pronounced reduction as compared to the calculated
value for pure Fe ($D = 2.73$ meV nm$^2$) but it remains still above
the experimental value for alloys with 25\% Al,
see Fig.~\ref{f_feal_dis}.

\begin{figure}[htb]
\begin{center}
\includegraphics[width=0.85\columnwidth]{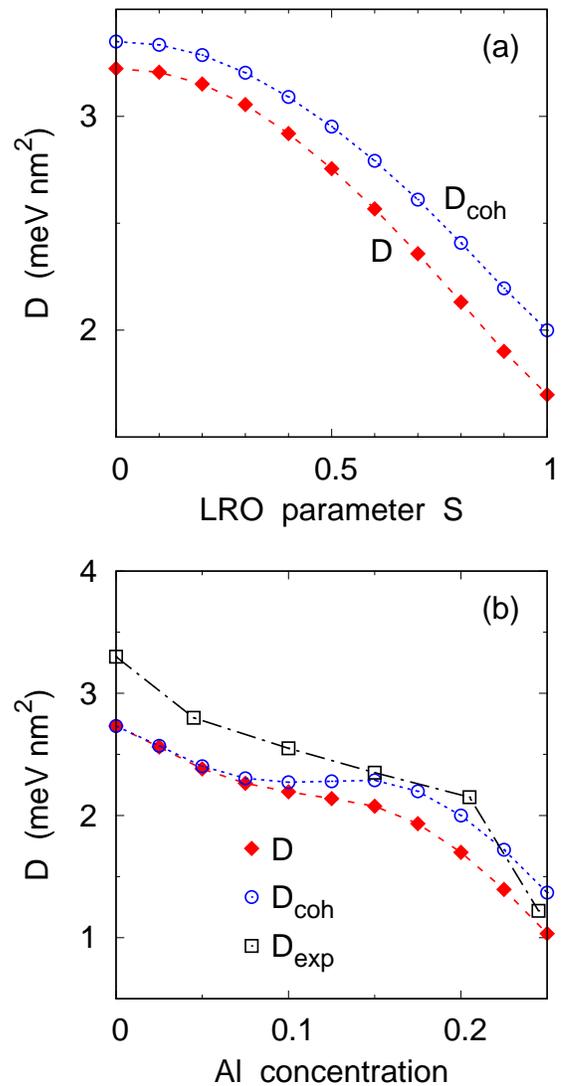}
\end{center}
\caption{
The spin-wave stiffness obtained from Eq.~(\ref{eq_dmn})
($D$, solid diamonds) and its coherent part ($D_\mathrm{coh}$,
open circles) in Fe-Al alloys with $B2$ atomic LRO:
(a) in Fe$_{0.8}$Al$_{0.2}$ alloy as functions of the LRO
parameter $S$, and 
(b) in Fe-Al alloys with maximum $B2$ LRO as functions of Al
concentration together with experimental values
\cite{r_1966_as, r_1964_hhl} ($D_\mathrm{exp}$, open boxes).
\label{f_feal_lro}}
\end{figure}

In order to assess the effect of the $B2$ atomic order on the
spin-wave stiffness, we have performed calculations for bcc Fe-Al
alloys with $B2$ atomic long-range order (LRO). 
The structure of an Fe$_{1-c}$Al$_c$ alloy thus contains two
simple cubic sublattices with respective compositions given by
Fe$_{1-c+u}$Al$_{c-u}$ and Fe$_{1-c-u}$Al$_{c+u}$, where $c$
denotes the global Al concentration and $u$ is an auxiliary 
concentration variable ($0 \le u \le c \le 0.25$).
The degree of the $B2$ LRO can be then quantified by a LRO parameter
$S = u/c$ ($0 \le S \le 1$); a completely random bcc alloy is given
by $S = 0$, whereas the value $S = 1$ refers to the maximum $B2$ LRO.
It should be noted that the developed formalism can be directly
extended to random systems with a few sublattices within the
single-site CPA, since the coherent potential function in
Eq.~(\ref{eq_gav}) is a site-diagonal matrix.
The calculated stiffness $D$ for the Fe$_{0.8}$Al$_{0.2}$ alloy as
a function of the LRO parameter $S$ is presented in 
Fig.~\ref{f_feal_lro}(a).
One can see a monotonic decrease of $D$ with increasing $S$;
the $B2$ LRO thus reduces the stiffness with a similar efficiency
as the $D0_3$ order.
Since the type and degree of the atomic order in experimentally
prepared Fe-Al samples is unknown, we have assumed the maximum
$B2$ LRO ($S = 1$) for each Al concentration and recalculated the
composition dependence of the stiffness.
As documented in Fig.~\ref{f_feal_lro}(b), the theoretical values of
$D$ reproduce now the measured data in a fair agreement, yielding
for 25\% Al the spin-wave stiffness as low as
$D \approx 1$ meV nm$^2$.

The theoretical results displayed in Fig.~\ref{f_feal_lro} include
also the coherent part of the spin-wave stiffness which enables one
to assess the role of the vertex corrections in this alloy system.
One finds that for the completely random bcc Fe$_{0.8}$Al$_{0.2}$
alloy, the vertex corrections are negative and their magnitude is
about 4\% of the total stiffness [Fig.~\ref{f_feal_lro}(a)].
The importance of the vertex corrections increases with increasing
Al content and $B2$ LRO; for Fe$_{0.75}$Al$_{0.25}$ and $S = 1$, the
relative magnitude of the incoherent part exceeds slightly 30\% of
the total $D$ [Fig.~\ref{f_feal_lro}(b)].
The negative sign and the appreciable magnitude of the vertex
corrections bring the theoretical values of $D$ into a better
agreement with the measured data.
A comparison of the role of the incoherent contribution to the
spin stiffness in the Fe-Al system to that in the Ni-Fe system
(Section~\ref{ss_nife}) reveals a striking analogy to a general
rule valid for residual electrical conductivities of random alloys
\cite{r_1986_sbs}: the vertex corrections are quite small in alloys
of transition metals (with dominating $d$ character of electron
states at the Fermi energy), but become significant in alloys
involving noble and simple metals.

\begin{table}[htb]
\caption{Local quantities of Fe atoms in three different phases (bcc,
$D0_3$, $B2$) of Fe$_{75}$Al$_{25}$ alloy: the average number of the
nearest-neighbor Fe atoms $N_\mathrm{Fe}$, the relative occurrence
$w$ of the particular Fe-atom position with respect to all Fe atoms,
the local magnetic moment $m$, and the on-site exchange parameter
$J^0$.
For the ordered phases, different sublattices occupied by Fe atoms
are denoted in the parentheses.
\label{t_fe75al}}
\begin{ruledtabular}
\begin{tabular}{lcccc}
phase & $N_\mathrm{Fe}$ & $w$ & $m$ ($\mu_\mathrm{B}$) & $J^0$ (mRy) \\
\hline
bcc & 6 & 1 & 2.19 & 12.6 \\
$D0_3$ (A,C) & 4 & 2/3 & 1.85 & \phantom{1}8.0 \\
$D0_3$ (B) & 8 & 1/3 & 2.37 & 17.3 \\
$B2$ (A) & 4 & 2/3 & 1.74 & \phantom{1}5.6 \\
$B2$ (B) & 8 & 1/3 & 2.49 & 14.9 \\
\end{tabular}
\end{ruledtabular}
\end{table}

A detailed microscopic explanation of the spin-wave softening
due to the atomic ordering in Fe-Al alloys goes beyond the scope of
the present work; nevertheless, its possible origin can be estimated
from an inspection of various site-resolved quantities of iron atoms.
A brief list of such quantities is presented in Table~\ref{t_fe75al}
for the Fe$_{75}$Al$_{25}$ alloy in three different phases:
the random bcc phase, the ordered $D0_3$ structure, and the alloy
with the maximum degree of $B2$ LRO.
Note that the $D0_3$ structure contains four fcc sublattices A, B, C,
and D, occupied by Fe atoms (A, B, C) and by Al atoms (D), whereby
the sublattices A and C are mutually equivalent.
The $B2$ phase with $S = 1$ consists of two simple cubic sublattices
A and B, with the former occupied solely by Fe atoms while the
chemical composition of the latter is Fe$_{50}$Al$_{50}$.
The average numbers $N_\mathrm{Fe}$ of nearest-neighbor Fe atoms of
central Fe atoms on different sublattices are displayed in
Table~\ref{t_fe75al} together with their local magnetic moments $m$
and relative occurrences $w$; the latter quantity is defined with
respect to all Fe atoms in the system, i.e., $w$ denotes the
probability that a randomly chosen Fe atom of the alloy occupies the
given sublattice (or any of sublattices A and C in the $D0_3$ phase). 
The ordering tendencies in the Fe-Al system reduce generally the
average number of Fe-Fe nearest neighbors; this reduction is
accompanied by a decrease of the local magnetic moments of Fe atoms
on the A (and C) sublattice in both ordered alloys, that is only
partly compensated by an increase of the Fe moments on the B
sublattice, see Table~\ref{t_fe75al}.
The ordering induces an even stronger decrease of an on-site exchange
parameter $J^0$ for Fe atoms on the A (and C) sublattice.
The on-site exchange parameter is defined in terms of the pair
exchange interactions as $J^0_\mathbf{R} = \sum_{\mathbf{R}'}
J_{\mathbf{R}\mathbf{R}'}$, it reflects the exchange field
experienced by the local magnetic moment at site $\mathbf{R}$, and it
is evaluated easily from the on-site blocks of the Green's functions
by using a well-known sum rule \cite{r_1987_lka, r_2006_tkd}.
The ordering-induced magnon softening can thus be ascribed to the
weakly coupled local moments of Fe atoms featured by a reduced number
of Fe nearest neighbors.
The validity of this conclusion is probably not confined only to the
studied Fe-Al system, but can also be extended to other iron-rich
alloys with $p$ elements (e.g., Fe-Si, Fe-Ga) where similar ordering
tendencies are encountered as well \cite{r_1986_tbm}. 

%\clearpage

\section{Conclusions\label{s_conc}}

In this work, a formulation of the spin-wave stiffness tensor of
itinerant ferromagnetic systems has been worked out by employing the
concepts and techniques used currently in the theory of electron
transport.
Application of the developed formalism to clean crystals allows
one to overcome convergence problems inherent to real-space lattice
summations involving pair exchange interactions.
The derived formulas can easily be combined with the CPA for an
efficient treatment of substitutionally disordered alloys, which
enables one to include the vertex corrections, often neglected in
existing calculations of the pair exchange interactions.

The first results of an implementation within the LSDA reproduced
successfully previous results of other authors for transition-metal
systems; in particular the decreasing trend of the spin-wave
stiffness with increasing Fe content in fcc Ni-Fe random alloys was
related to the concentration trend of the alloy magnetization.
For bcc Fe-Al random alloys, a strong sensitivity of the
spin-wave stiffness to the atomic order was proved indispensable
for a correct reproduction of the measured concentration dependence
by the calculations.
The two alloy systems studied represent two opposite cases from
a viewpoint of the vertex corrections: the latter are negligibly
small in the Ni-Fe alloys, but appreciable in the Fe-Al alloys.
These results follow thus similar findings obtained in the theory
of electron transport. 

The developed approach to the spin-wave stiffness tensor has been
presented in the TB-LMTO method but it can obviously be implemented
in the KKR multiple-scattering theory as well. 
An open question remains its possible generalization within a
relativistic theory of exchange interactions \cite{r_1992_al,
r_2003_usp} and micromagnetic parameters \cite{r_2014_fbm_c,
r_2019_mpe}; this has to be explored in the future.

\begin{acknowledgments}
This work was supported financially by the Czech Science
Foundation (Grant No.\ 18-07172S).
\end{acknowledgments}

%\bibliography{b_}

\begin{thebibliography}{60}%
\makeatletter
\providecommand \@ifxundefined [1]{%
 \@ifx{#1\undefined}
}%
\providecommand \@ifnum [1]{%
 \ifnum #1\expandafter \@firstoftwo
 \else \expandafter \@secondoftwo
 \fi
}%
\providecommand \@ifx [1]{%
 \ifx #1\expandafter \@firstoftwo
 \else \expandafter \@secondoftwo
 \fi
}%
\providecommand \natexlab [1]{#1}%
\providecommand \enquote  [1]{``#1''}%
\providecommand \bibnamefont  [1]{#1}%
\providecommand \bibfnamefont [1]{#1}%
\providecommand \citenamefont [1]{#1}%
\providecommand \href@noop [0]{\@secondoftwo}%
\providecommand \href [0]{\begingroup \@sanitize@url \@href}%
\providecommand \@href[1]{\@@startlink{#1}\@@href}%
\providecommand \@@href[1]{\endgroup#1\@@endlink}%
\providecommand \@sanitize@url [0]{\catcode `\\12\catcode `\$12\catcode
  `\&12\catcode `\#12\catcode `\^12\catcode `\_12\catcode `\%12\relax}%
\providecommand \@@startlink[1]{}%
\providecommand \@@endlink[0]{}%
\providecommand \url  [0]{\begingroup\@sanitize@url \@url }%
\providecommand \@url [1]{\endgroup\@href {#1}{\urlprefix }}%
\providecommand \urlprefix  [0]{URL }%
\providecommand \Eprint [0]{\href }%
\providecommand \doibase [0]{http://dx.doi.org/}%
\providecommand \selectlanguage [0]{\@gobble}%
\providecommand \bibinfo  [0]{\@secondoftwo}%
\providecommand \bibfield  [0]{\@secondoftwo}%
\providecommand \translation [1]{[#1]}%
\providecommand \BibitemOpen [0]{}%
\providecommand \bibitemStop [0]{}%
\providecommand \bibitemNoStop [0]{.\EOS\space}%
\providecommand \EOS [0]{\spacefactor3000\relax}%
\providecommand \BibitemShut  [1]{\csname bibitem#1\endcsname}%
\let\auto@bib@innerbib\@empty
%</preamble>
\bibitem [{\citenamefont {Moreno}\ \emph {et~al.}(2016)\citenamefont {Moreno},
  \citenamefont {Evans}, \citenamefont {Khmelevskyi}, \citenamefont
  {Mu{\~n}oz}, \citenamefont {Chantrell},\ and\ \citenamefont
  {Chubykalo-Fesenko}}]{r_2016_mek}%
  \BibitemOpen
  \bibfield  {author} {\bibinfo {author} {\bibfnamefont {R.}~\bibnamefont
  {Moreno}}, \bibinfo {author} {\bibfnamefont {R.~F.~L.}\ \bibnamefont
  {Evans}}, \bibinfo {author} {\bibfnamefont {S.}~\bibnamefont {Khmelevskyi}},
  \bibinfo {author} {\bibfnamefont {M.~C.}\ \bibnamefont {Mu{\~n}oz}}, \bibinfo
  {author} {\bibfnamefont {R.~W.}\ \bibnamefont {Chantrell}}, \ and\ \bibinfo
  {author} {\bibfnamefont {O.}~\bibnamefont {Chubykalo-Fesenko}},\ }\href@noop
  {} {\bibfield  {journal} {\bibinfo  {journal} {Phys. Rev. B}\ }\textbf
  {\bibinfo {volume} {94}},\ \bibinfo {pages} {104433} (\bibinfo {year}
  {2016})}\BibitemShut {NoStop}%
\bibitem [{\citenamefont {N\v{e}mec}\ \emph {et~al.}(2013)\citenamefont
  {N\v{e}mec}, \citenamefont {Nov\'ak}, \citenamefont {Tesa\v{r}ov\'a},
  \citenamefont {Rozkotov\'a}, \citenamefont {Reichlov\'a}, \citenamefont
  {Butkovi\v{c}ov\'a}, \citenamefont {Troj\'anek}, \citenamefont
  {Olejn\'{\i}k}, \citenamefont {Mal\'y}, \citenamefont {Campion},
  \citenamefont {Gallagher}, \citenamefont {Sinova},\ and\ \citenamefont
  {Jungwirth}}]{r_2013_nnt}%
  \BibitemOpen
  \bibfield  {author} {\bibinfo {author} {\bibfnamefont {P.}~\bibnamefont
  {N\v{e}mec}}, \bibinfo {author} {\bibfnamefont {V.}~\bibnamefont {Nov\'ak}},
  \bibinfo {author} {\bibfnamefont {N.}~\bibnamefont {Tesa\v{r}ov\'a}},
  \bibinfo {author} {\bibfnamefont {E.}~\bibnamefont {Rozkotov\'a}}, \bibinfo
  {author} {\bibfnamefont {H.}~\bibnamefont {Reichlov\'a}}, \bibinfo {author}
  {\bibfnamefont {D.}~\bibnamefont {Butkovi\v{c}ov\'a}}, \bibinfo {author}
  {\bibfnamefont {F.}~\bibnamefont {Troj\'anek}}, \bibinfo {author}
  {\bibfnamefont {K.}~\bibnamefont {Olejn\'{\i}k}}, \bibinfo {author}
  {\bibfnamefont {P.}~\bibnamefont {Mal\'y}}, \bibinfo {author} {\bibfnamefont
  {R.~P.}\ \bibnamefont {Campion}}, \bibinfo {author} {\bibfnamefont {B.~L.}\
  \bibnamefont {Gallagher}}, \bibinfo {author} {\bibfnamefont {J.}~\bibnamefont
  {Sinova}}, \ and\ \bibinfo {author} {\bibfnamefont {T.}~\bibnamefont
  {Jungwirth}},\ }\href@noop {} {\bibfield  {journal} {\bibinfo  {journal}
  {Nat. Commun.}\ }\textbf {\bibinfo {volume} {4}},\ \bibinfo {pages} {1422}
  (\bibinfo {year} {2013})}\BibitemShut {NoStop}%
\bibitem [{\citenamefont {Callaway}\ \emph {et~al.}(1981)\citenamefont
  {Callaway}, \citenamefont {Wang},\ and\ \citenamefont
  {Laurent}}]{r_1981_cwl}%
  \BibitemOpen
  \bibfield  {author} {\bibinfo {author} {\bibfnamefont {J.}~\bibnamefont
  {Callaway}}, \bibinfo {author} {\bibfnamefont {C.~S.}\ \bibnamefont {Wang}},
  \ and\ \bibinfo {author} {\bibfnamefont {D.~G.}\ \bibnamefont {Laurent}},\
  }\href@noop {} {\bibfield  {journal} {\bibinfo  {journal} {Phys. Rev. B}\
  }\textbf {\bibinfo {volume} {24}},\ \bibinfo {pages} {6491} (\bibinfo {year}
  {1981})}\BibitemShut {NoStop}%
\bibitem [{\citenamefont {Sandratskii}(1998)}]{r_1998_lms}%
  \BibitemOpen
  \bibfield  {author} {\bibinfo {author} {\bibfnamefont {L.~M.}\ \bibnamefont
  {Sandratskii}},\ }\href@noop {} {\bibfield  {journal} {\bibinfo  {journal}
  {Adv. Phys.}\ }\textbf {\bibinfo {volume} {47}},\ \bibinfo {pages} {91}
  (\bibinfo {year} {1998})}\BibitemShut {NoStop}%
\bibitem [{\citenamefont {Rosengaard}\ and\ \citenamefont
  {Johansson}(1997)}]{r_1997_rj}%
  \BibitemOpen
  \bibfield  {author} {\bibinfo {author} {\bibfnamefont {N.~M.}\ \bibnamefont
  {Rosengaard}}\ and\ \bibinfo {author} {\bibfnamefont {B.}~\bibnamefont
  {Johansson}},\ }\href@noop {} {\bibfield  {journal} {\bibinfo  {journal}
  {Phys. Rev. B}\ }\textbf {\bibinfo {volume} {55}},\ \bibinfo {pages} {14975}
  (\bibinfo {year} {1997})}\BibitemShut {NoStop}%
\bibitem [{\citenamefont {Halilov}\ \emph {et~al.}(1998)\citenamefont
  {Halilov}, \citenamefont {Eschrig}, \citenamefont {Perlov},\ and\
  \citenamefont {Oppeneer}}]{r_1998_hep}%
  \BibitemOpen
  \bibfield  {author} {\bibinfo {author} {\bibfnamefont {S.~V.}\ \bibnamefont
  {Halilov}}, \bibinfo {author} {\bibfnamefont {H.}~\bibnamefont {Eschrig}},
  \bibinfo {author} {\bibfnamefont {A.~Y.}\ \bibnamefont {Perlov}}, \ and\
  \bibinfo {author} {\bibfnamefont {P.~M.}\ \bibnamefont {Oppeneer}},\
  }\href@noop {} {\bibfield  {journal} {\bibinfo  {journal} {Phys. Rev. B}\
  }\textbf {\bibinfo {volume} {58}},\ \bibinfo {pages} {293} (\bibinfo {year}
  {1998})}\BibitemShut {NoStop}%
\bibitem [{\citenamefont {Brown}\ \emph {et~al.}(1999)\citenamefont {Brown},
  \citenamefont {Nicholson}, \citenamefont {Wang},\ and\ \citenamefont
  {Schulthess}}]{r_1999_bnw}%
  \BibitemOpen
  \bibfield  {author} {\bibinfo {author} {\bibfnamefont {R.~H.}\ \bibnamefont
  {Brown}}, \bibinfo {author} {\bibfnamefont {D.~M.~C.}\ \bibnamefont
  {Nicholson}}, \bibinfo {author} {\bibfnamefont {X.}~\bibnamefont {Wang}}, \
  and\ \bibinfo {author} {\bibfnamefont {T.~C.}\ \bibnamefont {Schulthess}},\
  }\href@noop {} {\bibfield  {journal} {\bibinfo  {journal} {J. Appl. Phys.}\
  }\textbf {\bibinfo {volume} {85}},\ \bibinfo {pages} {4830} (\bibinfo {year}
  {1999})}\BibitemShut {NoStop}%
\bibitem [{\citenamefont {Liechtenstein}\ \emph {et~al.}(1984)\citenamefont
  {Liechtenstein}, \citenamefont {Katsnelson},\ and\ \citenamefont
  {Gubanov}}]{r_1984_lkg}%
  \BibitemOpen
  \bibfield  {author} {\bibinfo {author} {\bibfnamefont {A.~I.}\ \bibnamefont
  {Liechtenstein}}, \bibinfo {author} {\bibfnamefont {M.~I.}\ \bibnamefont
  {Katsnelson}}, \ and\ \bibinfo {author} {\bibfnamefont {V.~A.}\ \bibnamefont
  {Gubanov}},\ }\href@noop {} {\bibfield  {journal} {\bibinfo  {journal} {J.
  Phys. F: Met. Phys.}\ }\textbf {\bibinfo {volume} {14}},\ \bibinfo {pages}
  {L125} (\bibinfo {year} {1984})}\BibitemShut {NoStop}%
\bibitem [{\citenamefont {Liechtenstein}\ \emph {et~al.}(1987)\citenamefont
  {Liechtenstein}, \citenamefont {Katsnelson}, \citenamefont {Antropov},\ and\
  \citenamefont {Gubanov}}]{r_1987_lka}%
  \BibitemOpen
  \bibfield  {author} {\bibinfo {author} {\bibfnamefont {A.~I.}\ \bibnamefont
  {Liechtenstein}}, \bibinfo {author} {\bibfnamefont {M.~I.}\ \bibnamefont
  {Katsnelson}}, \bibinfo {author} {\bibfnamefont {V.~P.}\ \bibnamefont
  {Antropov}}, \ and\ \bibinfo {author} {\bibfnamefont {V.~A.}\ \bibnamefont
  {Gubanov}},\ }\href@noop {} {\bibfield  {journal} {\bibinfo  {journal} {J.
  Magn. Magn. Mater.}\ }\textbf {\bibinfo {volume} {67}},\ \bibinfo {pages}
  {65} (\bibinfo {year} {1987})}\BibitemShut {NoStop}%
\bibitem [{\citenamefont {Pajda}\ \emph {et~al.}(2001)\citenamefont {Pajda},
  \citenamefont {Kudrnovsk\'y}, \citenamefont {Turek}, \citenamefont {Drchal},\
  and\ \citenamefont {Bruno}}]{r_2001_pkt}%
  \BibitemOpen
  \bibfield  {author} {\bibinfo {author} {\bibfnamefont {M.}~\bibnamefont
  {Pajda}}, \bibinfo {author} {\bibfnamefont {J.}~\bibnamefont {Kudrnovsk\'y}},
  \bibinfo {author} {\bibfnamefont {I.}~\bibnamefont {Turek}}, \bibinfo
  {author} {\bibfnamefont {V.}~\bibnamefont {Drchal}}, \ and\ \bibinfo {author}
  {\bibfnamefont {P.}~\bibnamefont {Bruno}},\ }\href@noop {} {\bibfield
  {journal} {\bibinfo  {journal} {Phys. Rev. B}\ }\textbf {\bibinfo {volume}
  {64}},\ \bibinfo {pages} {174402} (\bibinfo {year} {2001})}\BibitemShut
  {NoStop}%
\bibitem [{\citenamefont {Bose}\ \emph {et~al.}(2010)\citenamefont {Bose},
  \citenamefont {Kudrnovsk\'y}, \citenamefont {Drchal},\ and\ \citenamefont
  {Turek}}]{r_2010_bkd}%
  \BibitemOpen
  \bibfield  {author} {\bibinfo {author} {\bibfnamefont {S.~K.}\ \bibnamefont
  {Bose}}, \bibinfo {author} {\bibfnamefont {J.}~\bibnamefont {Kudrnovsk\'y}},
  \bibinfo {author} {\bibfnamefont {V.}~\bibnamefont {Drchal}}, \ and\ \bibinfo
  {author} {\bibfnamefont {I.}~\bibnamefont {Turek}},\ }\href@noop {}
  {\bibfield  {journal} {\bibinfo  {journal} {Phys. Rev. B}\ }\textbf {\bibinfo
  {volume} {82}},\ \bibinfo {pages} {174402} (\bibinfo {year}
  {2010})}\BibitemShut {NoStop}%
\bibitem [{\citenamefont {\v{S}ipr}\ \emph {et~al.}(2019)\citenamefont
  {\v{S}ipr}, \citenamefont {Mankovsky},\ and\ \citenamefont
  {Ebert}}]{r_2019_sme}%
  \BibitemOpen
  \bibfield  {author} {\bibinfo {author} {\bibfnamefont {O.}~\bibnamefont
  {\v{S}ipr}}, \bibinfo {author} {\bibfnamefont {S.}~\bibnamefont {Mankovsky}},
  \ and\ \bibinfo {author} {\bibfnamefont {H.}~\bibnamefont {Ebert}},\
  }\href@noop {} {\bibfield  {journal} {\bibinfo  {journal} {Phys. Rev. B}\
  }\textbf {\bibinfo {volume} {100}},\ \bibinfo {pages} {024435} (\bibinfo
  {year} {2019})}\BibitemShut {NoStop}%
\bibitem [{\citenamefont {Mankovsky}\ \emph {et~al.}(2011)\citenamefont
  {Mankovsky}, \citenamefont {Fecher},\ and\ \citenamefont
  {Ebert}}]{r_2011_mfe}%
  \BibitemOpen
  \bibfield  {author} {\bibinfo {author} {\bibfnamefont {S.}~\bibnamefont
  {Mankovsky}}, \bibinfo {author} {\bibfnamefont {G.~H.}\ \bibnamefont
  {Fecher}}, \ and\ \bibinfo {author} {\bibfnamefont {H.}~\bibnamefont
  {Ebert}},\ }\href@noop {} {\bibfield  {journal} {\bibinfo  {journal} {Phys.
  Rev. B}\ }\textbf {\bibinfo {volume} {83}},\ \bibinfo {pages} {144401}
  (\bibinfo {year} {2011})}\BibitemShut {NoStop}%
\bibitem [{\citenamefont {Turek}\ \emph {et~al.}(2006)\citenamefont {Turek},
  \citenamefont {Kudrnovsk\'y}, \citenamefont {Drchal},\ and\ \citenamefont
  {Bruno}}]{r_2006_tkd}%
  \BibitemOpen
  \bibfield  {author} {\bibinfo {author} {\bibfnamefont {I.}~\bibnamefont
  {Turek}}, \bibinfo {author} {\bibfnamefont {J.}~\bibnamefont {Kudrnovsk\'y}},
  \bibinfo {author} {\bibfnamefont {V.}~\bibnamefont {Drchal}}, \ and\ \bibinfo
  {author} {\bibfnamefont {P.}~\bibnamefont {Bruno}},\ }\href@noop {}
  {\bibfield  {journal} {\bibinfo  {journal} {Philos. Mag.}\ }\textbf {\bibinfo
  {volume} {86}},\ \bibinfo {pages} {1713} (\bibinfo {year}
  {2006})}\BibitemShut {NoStop}%
\bibitem [{\citenamefont {Kudrnovsk\'y}\ \emph {et~al.}(2008)\citenamefont
  {Kudrnovsk\'y}, \citenamefont {Drchal},\ and\ \citenamefont
  {Bruno}}]{r_2008_kdb}%
  \BibitemOpen
  \bibfield  {author} {\bibinfo {author} {\bibfnamefont {J.}~\bibnamefont
  {Kudrnovsk\'y}}, \bibinfo {author} {\bibfnamefont {V.}~\bibnamefont
  {Drchal}}, \ and\ \bibinfo {author} {\bibfnamefont {P.}~\bibnamefont
  {Bruno}},\ }\href@noop {} {\bibfield  {journal} {\bibinfo  {journal} {Phys.
  Rev. B}\ }\textbf {\bibinfo {volume} {77}},\ \bibinfo {pages} {224422}
  (\bibinfo {year} {2008})}\BibitemShut {NoStop}%
\bibitem [{\citenamefont {Ruban}\ \emph {et~al.}(2005)\citenamefont {Ruban},
  \citenamefont {Katsnelson}, \citenamefont {Olovsson}, \citenamefont {Simak},\
  and\ \citenamefont {Abrikosov}}]{r_2005_rko}%
  \BibitemOpen
  \bibfield  {author} {\bibinfo {author} {\bibfnamefont {A.~V.}\ \bibnamefont
  {Ruban}}, \bibinfo {author} {\bibfnamefont {M.~I.}\ \bibnamefont
  {Katsnelson}}, \bibinfo {author} {\bibfnamefont {W.}~\bibnamefont
  {Olovsson}}, \bibinfo {author} {\bibfnamefont {S.~I.}\ \bibnamefont {Simak}},
  \ and\ \bibinfo {author} {\bibfnamefont {I.~A.}\ \bibnamefont {Abrikosov}},\
  }\href@noop {} {\bibfield  {journal} {\bibinfo  {journal} {Phys. Rev. B}\
  }\textbf {\bibinfo {volume} {71}},\ \bibinfo {pages} {054402} (\bibinfo
  {year} {2005})}\BibitemShut {NoStop}%
\bibitem [{\citenamefont {Velick\'y}(1969)}]{r_1969_bv}%
  \BibitemOpen
  \bibfield  {author} {\bibinfo {author} {\bibfnamefont {B.}~\bibnamefont
  {Velick\'y}},\ }\href@noop {} {\bibfield  {journal} {\bibinfo  {journal}
  {Phys. Rev.}\ }\textbf {\bibinfo {volume} {184}},\ \bibinfo {pages} {614}
  (\bibinfo {year} {1969})}\BibitemShut {NoStop}%
\bibitem [{\citenamefont {Bruno}\ \emph {et~al.}(1996)\citenamefont {Bruno},
  \citenamefont {Kudrnovsk\'y}, \citenamefont {Drchal},\ and\ \citenamefont
  {Turek}}]{r_1996_bkd}%
  \BibitemOpen
  \bibfield  {author} {\bibinfo {author} {\bibfnamefont {P.}~\bibnamefont
  {Bruno}}, \bibinfo {author} {\bibfnamefont {J.}~\bibnamefont {Kudrnovsk\'y}},
  \bibinfo {author} {\bibfnamefont {V.}~\bibnamefont {Drchal}}, \ and\ \bibinfo
  {author} {\bibfnamefont {I.}~\bibnamefont {Turek}},\ }\href@noop {}
  {\bibfield  {journal} {\bibinfo  {journal} {Phys. Rev. Lett.}\ }\textbf
  {\bibinfo {volume} {76}},\ \bibinfo {pages} {4254} (\bibinfo {year}
  {1996})}\BibitemShut {NoStop}%
\bibitem [{\citenamefont {Buczek}\ \emph {et~al.}(2018)\citenamefont {Buczek},
  \citenamefont {Thomas}, \citenamefont {Marmodoro}, \citenamefont {Buczek},
  \citenamefont {Zubizarreta}, \citenamefont {Hoffmann}, \citenamefont
  {Balashov}, \citenamefont {Wulfhekel}, \citenamefont {Zakeri},\ and\
  \citenamefont {Ernst}}]{r_2018_btm}%
  \BibitemOpen
  \bibfield  {author} {\bibinfo {author} {\bibfnamefont {P.}~\bibnamefont
  {Buczek}}, \bibinfo {author} {\bibfnamefont {S.}~\bibnamefont {Thomas}},
  \bibinfo {author} {\bibfnamefont {A.}~\bibnamefont {Marmodoro}}, \bibinfo
  {author} {\bibfnamefont {N.}~\bibnamefont {Buczek}}, \bibinfo {author}
  {\bibfnamefont {X.}~\bibnamefont {Zubizarreta}}, \bibinfo {author}
  {\bibfnamefont {M.}~\bibnamefont {Hoffmann}}, \bibinfo {author}
  {\bibfnamefont {T.}~\bibnamefont {Balashov}}, \bibinfo {author}
  {\bibfnamefont {W.}~\bibnamefont {Wulfhekel}}, \bibinfo {author}
  {\bibfnamefont {K.}~\bibnamefont {Zakeri}}, \ and\ \bibinfo {author}
  {\bibfnamefont {A.}~\bibnamefont {Ernst}},\ }\href@noop {} {\bibfield
  {journal} {\bibinfo  {journal} {J. Phys.: Condens. Matter}\ }\textbf
  {\bibinfo {volume} {30}},\ \bibinfo {pages} {423001} (\bibinfo {year}
  {2018})}\BibitemShut {NoStop}%
\bibitem [{\citenamefont {Bouzerar}\ and\ \citenamefont
  {Bruno}(2002)}]{r_2002_bb}%
  \BibitemOpen
  \bibfield  {author} {\bibinfo {author} {\bibfnamefont {G.}~\bibnamefont
  {Bouzerar}}\ and\ \bibinfo {author} {\bibfnamefont {P.}~\bibnamefont
  {Bruno}},\ }\href@noop {} {\bibfield  {journal} {\bibinfo  {journal} {Phys.
  Rev. B}\ }\textbf {\bibinfo {volume} {66}},\ \bibinfo {pages} {014410}
  (\bibinfo {year} {2002})}\BibitemShut {NoStop}%
\bibitem [{\citenamefont {Buczek}\ \emph {et~al.}(2016)\citenamefont {Buczek},
  \citenamefont {Sandratskii}, \citenamefont {Buczek}, \citenamefont {Thomas},
  \citenamefont {Vignale},\ and\ \citenamefont {Ernst}}]{r_2016_bsb}%
  \BibitemOpen
  \bibfield  {author} {\bibinfo {author} {\bibfnamefont {P.}~\bibnamefont
  {Buczek}}, \bibinfo {author} {\bibfnamefont {L.~M.}\ \bibnamefont
  {Sandratskii}}, \bibinfo {author} {\bibfnamefont {N.}~\bibnamefont {Buczek}},
  \bibinfo {author} {\bibfnamefont {S.}~\bibnamefont {Thomas}}, \bibinfo
  {author} {\bibfnamefont {G.}~\bibnamefont {Vignale}}, \ and\ \bibinfo
  {author} {\bibfnamefont {A.}~\bibnamefont {Ernst}},\ }\href@noop {}
  {\bibfield  {journal} {\bibinfo  {journal} {Phys. Rev. B}\ }\textbf {\bibinfo
  {volume} {94}},\ \bibinfo {pages} {054407} (\bibinfo {year}
  {2016})}\BibitemShut {NoStop}%
\bibitem [{\citenamefont {Skubic}\ \emph {et~al.}(2008)\citenamefont {Skubic},
  \citenamefont {Hellsvik}, \citenamefont {Nordstr{\"o}m},\ and\ \citenamefont
  {Eriksson}}]{r_2008_shn}%
  \BibitemOpen
  \bibfield  {author} {\bibinfo {author} {\bibfnamefont {B.}~\bibnamefont
  {Skubic}}, \bibinfo {author} {\bibfnamefont {J.}~\bibnamefont {Hellsvik}},
  \bibinfo {author} {\bibfnamefont {L.}~\bibnamefont {Nordstr{\"o}m}}, \ and\
  \bibinfo {author} {\bibfnamefont {O.}~\bibnamefont {Eriksson}},\ }\href@noop
  {} {\bibfield  {journal} {\bibinfo  {journal} {J. Phys.: Condens. Matter}\
  }\textbf {\bibinfo {volume} {20}},\ \bibinfo {pages} {315203} (\bibinfo
  {year} {2008})}\BibitemShut {NoStop}%
\bibitem [{\citenamefont {Bouzerar}(2007)}]{r_2007_gb}%
  \BibitemOpen
  \bibfield  {author} {\bibinfo {author} {\bibfnamefont {G.}~\bibnamefont
  {Bouzerar}},\ }\href@noop {} {\bibfield  {journal} {\bibinfo  {journal}
  {Europhys. Lett.}\ }\textbf {\bibinfo {volume} {79}},\ \bibinfo {pages}
  {57007} (\bibinfo {year} {2007})}\BibitemShut {NoStop}%
\bibitem [{\citenamefont {Turek}\ \emph {et~al.}(2016)\citenamefont {Turek},
  \citenamefont {Kudrnovsk\'y},\ and\ \citenamefont {Drchal}}]{r_2016_tkd}%
  \BibitemOpen
  \bibfield  {author} {\bibinfo {author} {\bibfnamefont {I.}~\bibnamefont
  {Turek}}, \bibinfo {author} {\bibfnamefont {J.}~\bibnamefont {Kudrnovsk\'y}},
  \ and\ \bibinfo {author} {\bibfnamefont {V.}~\bibnamefont {Drchal}},\
  }\href@noop {} {\bibfield  {journal} {\bibinfo  {journal} {Phys. Rev. B}\
  }\textbf {\bibinfo {volume} {94}},\ \bibinfo {pages} {174447} (\bibinfo
  {year} {2016})}\BibitemShut {NoStop}%
\bibitem [{\citenamefont {Freimuth}\ \emph {et~al.}(2014)\citenamefont
  {Freimuth}, \citenamefont {Bl{\"u}gel},\ and\ \citenamefont
  {Mokrousov}}]{r_2014_fbm_c}%
  \BibitemOpen
  \bibfield  {author} {\bibinfo {author} {\bibfnamefont {F.}~\bibnamefont
  {Freimuth}}, \bibinfo {author} {\bibfnamefont {S.}~\bibnamefont
  {Bl{\"u}gel}}, \ and\ \bibinfo {author} {\bibfnamefont {Y.}~\bibnamefont
  {Mokrousov}},\ }\href@noop {} {\bibfield  {journal} {\bibinfo  {journal} {J.
  Phys.: Condens. Matter}\ }\textbf {\bibinfo {volume} {26}},\ \bibinfo {pages}
  {104202} (\bibinfo {year} {2014})}\BibitemShut {NoStop}%
\bibitem [{\citenamefont {Mankovsky}\ \emph {et~al.}(2019)\citenamefont
  {Mankovsky}, \citenamefont {Polesya},\ and\ \citenamefont
  {Ebert}}]{r_2019_mpe}%
  \BibitemOpen
  \bibfield  {author} {\bibinfo {author} {\bibfnamefont {S.}~\bibnamefont
  {Mankovsky}}, \bibinfo {author} {\bibfnamefont {S.}~\bibnamefont {Polesya}},
  \ and\ \bibinfo {author} {\bibfnamefont {H.}~\bibnamefont {Ebert}},\
  }\href@noop {} {\bibfield  {journal} {\bibinfo  {journal} {Phys. Rev. B}\
  }\textbf {\bibinfo {volume} {99}},\ \bibinfo {pages} {104427} (\bibinfo
  {year} {2019})}\BibitemShut {NoStop}%
\bibitem [{\citenamefont {Nagaosa}\ \emph {et~al.}(2010)\citenamefont
  {Nagaosa}, \citenamefont {Sinova}, \citenamefont {Onoda}, \citenamefont
  {MacDonald},\ and\ \citenamefont {Ong}}]{r_2010_nso}%
  \BibitemOpen
  \bibfield  {author} {\bibinfo {author} {\bibfnamefont {N.}~\bibnamefont
  {Nagaosa}}, \bibinfo {author} {\bibfnamefont {J.}~\bibnamefont {Sinova}},
  \bibinfo {author} {\bibfnamefont {S.}~\bibnamefont {Onoda}}, \bibinfo
  {author} {\bibfnamefont {A.~H.}\ \bibnamefont {MacDonald}}, \ and\ \bibinfo
  {author} {\bibfnamefont {N.~P.}\ \bibnamefont {Ong}},\ }\href@noop {}
  {\bibfield  {journal} {\bibinfo  {journal} {Rev. Mod. Phys.}\ }\textbf
  {\bibinfo {volume} {82}},\ \bibinfo {pages} {1539} (\bibinfo {year}
  {2010})}\BibitemShut {NoStop}%
\bibitem [{\citenamefont {\v{S}ipr}\ \emph {et~al.}(2020)\citenamefont
  {\v{S}ipr}, \citenamefont {Mankovsky},\ and\ \citenamefont
  {Ebert}}]{r_2020_sme}%
  \BibitemOpen
  \bibfield  {author} {\bibinfo {author} {\bibfnamefont {O.}~\bibnamefont
  {\v{S}ipr}}, \bibinfo {author} {\bibfnamefont {S.}~\bibnamefont {Mankovsky}},
  \ and\ \bibinfo {author} {\bibfnamefont {H.}~\bibnamefont {Ebert}},\
  }\href@noop {} {}\bibinfo {howpublished} {arXiv:2001.02558} (\bibinfo {year}
  {2020})\BibitemShut {NoStop}%
\bibitem [{\citenamefont {Kudrnovsk\'y}\ \emph {et~al.}(2014)\citenamefont
  {Kudrnovsk\'y}, \citenamefont {Drchal}, \citenamefont {Bergqvist},
  \citenamefont {Rusz}, \citenamefont {Turek}, \citenamefont {\'Ujfalussy},\
  and\ \citenamefont {Vincze}}]{r_2014_kdb}%
  \BibitemOpen
  \bibfield  {author} {\bibinfo {author} {\bibfnamefont {J.}~\bibnamefont
  {Kudrnovsk\'y}}, \bibinfo {author} {\bibfnamefont {V.}~\bibnamefont
  {Drchal}}, \bibinfo {author} {\bibfnamefont {L.}~\bibnamefont {Bergqvist}},
  \bibinfo {author} {\bibfnamefont {J.}~\bibnamefont {Rusz}}, \bibinfo {author}
  {\bibfnamefont {I.}~\bibnamefont {Turek}}, \bibinfo {author} {\bibfnamefont
  {B.}~\bibnamefont {\'Ujfalussy}}, \ and\ \bibinfo {author} {\bibfnamefont
  {I.}~\bibnamefont {Vincze}},\ }\href@noop {} {\bibfield  {journal} {\bibinfo
  {journal} {Phys. Rev. B}\ }\textbf {\bibinfo {volume} {90}},\ \bibinfo
  {pages} {134408} (\bibinfo {year} {2014})}\BibitemShut {NoStop}%
\bibitem [{\citenamefont {Zarestky}\ \emph {et~al.}(2005)\citenamefont
  {Zarestky}, \citenamefont {Garlea}, \citenamefont {Lograsso}, \citenamefont
  {Schlagel},\ and\ \citenamefont {Stassis}}]{r_2005_zgl}%
  \BibitemOpen
  \bibfield  {author} {\bibinfo {author} {\bibfnamefont {J.~L.}\ \bibnamefont
  {Zarestky}}, \bibinfo {author} {\bibfnamefont {V.~O.}\ \bibnamefont
  {Garlea}}, \bibinfo {author} {\bibfnamefont {T.~A.}\ \bibnamefont
  {Lograsso}}, \bibinfo {author} {\bibfnamefont {D.~L.}\ \bibnamefont
  {Schlagel}}, \ and\ \bibinfo {author} {\bibfnamefont {C.}~\bibnamefont
  {Stassis}},\ }\href@noop {} {\bibfield  {journal} {\bibinfo  {journal} {Phys.
  Rev. B}\ }\textbf {\bibinfo {volume} {72}},\ \bibinfo {pages} {180408(R)}
  (\bibinfo {year} {2005})}\BibitemShut {NoStop}%
\bibitem [{\citenamefont {Zhao}\ \emph {et~al.}(2006)\citenamefont {Zhao},
  \citenamefont {Cullen}, \citenamefont {Wuttig}, \citenamefont {Kang},
  \citenamefont {Lynn}, \citenamefont {Lograsso},\ and\ \citenamefont
  {Moze}}]{r_2006_zcw}%
  \BibitemOpen
  \bibfield  {author} {\bibinfo {author} {\bibfnamefont {P.}~\bibnamefont
  {Zhao}}, \bibinfo {author} {\bibfnamefont {J.}~\bibnamefont {Cullen}},
  \bibinfo {author} {\bibfnamefont {M.}~\bibnamefont {Wuttig}}, \bibinfo
  {author} {\bibfnamefont {H.~J.}\ \bibnamefont {Kang}}, \bibinfo {author}
  {\bibfnamefont {J.~W.}\ \bibnamefont {Lynn}}, \bibinfo {author}
  {\bibfnamefont {T.~A.}\ \bibnamefont {Lograsso}}, \ and\ \bibinfo {author}
  {\bibfnamefont {O.}~\bibnamefont {Moze}},\ }\href@noop {} {\bibfield
  {journal} {\bibinfo  {journal} {J. Appl. Phys.}\ }\textbf {\bibinfo {volume}
  {99}},\ \bibinfo {pages} {08R101} (\bibinfo {year} {2006})}\BibitemShut
  {NoStop}%
\bibitem [{\citenamefont {Zarestky}\ \emph {et~al.}(2007)\citenamefont
  {Zarestky}, \citenamefont {Moze}, \citenamefont {Lynn}, \citenamefont {Chen},
  \citenamefont {Lograsso},\ and\ \citenamefont {Schlagel}}]{r_2007_zml}%
  \BibitemOpen
  \bibfield  {author} {\bibinfo {author} {\bibfnamefont {J.~L.}\ \bibnamefont
  {Zarestky}}, \bibinfo {author} {\bibfnamefont {O.}~\bibnamefont {Moze}},
  \bibinfo {author} {\bibfnamefont {J.~W.}\ \bibnamefont {Lynn}}, \bibinfo
  {author} {\bibfnamefont {Y.}~\bibnamefont {Chen}}, \bibinfo {author}
  {\bibfnamefont {T.~A.}\ \bibnamefont {Lograsso}}, \ and\ \bibinfo {author}
  {\bibfnamefont {D.~L.}\ \bibnamefont {Schlagel}},\ }\href@noop {} {\bibfield
  {journal} {\bibinfo  {journal} {Phys. Rev. B}\ }\textbf {\bibinfo {volume}
  {75}},\ \bibinfo {pages} {052406} (\bibinfo {year} {2007})}\BibitemShut
  {NoStop}%
\bibitem [{\citenamefont {Andersen}(1975)}]{r_1975_oka}%
  \BibitemOpen
  \bibfield  {author} {\bibinfo {author} {\bibfnamefont {O.~K.}\ \bibnamefont
  {Andersen}},\ }\href@noop {} {\bibfield  {journal} {\bibinfo  {journal}
  {Phys. Rev. B}\ }\textbf {\bibinfo {volume} {12}},\ \bibinfo {pages} {3060}
  (\bibinfo {year} {1975})}\BibitemShut {NoStop}%
\bibitem [{\citenamefont {Gunnarsson}\ \emph {et~al.}(1983)\citenamefont
  {Gunnarsson}, \citenamefont {Jepsen},\ and\ \citenamefont
  {Andersen}}]{r_1983_gja}%
  \BibitemOpen
  \bibfield  {author} {\bibinfo {author} {\bibfnamefont {O.}~\bibnamefont
  {Gunnarsson}}, \bibinfo {author} {\bibfnamefont {O.}~\bibnamefont {Jepsen}},
  \ and\ \bibinfo {author} {\bibfnamefont {O.~K.}\ \bibnamefont {Andersen}},\
  }\href@noop {} {\bibfield  {journal} {\bibinfo  {journal} {Phys. Rev. B}\
  }\textbf {\bibinfo {volume} {27}},\ \bibinfo {pages} {7144} (\bibinfo {year}
  {1983})}\BibitemShut {NoStop}%
\bibitem [{\citenamefont {Turek}\ \emph {et~al.}(1997)\citenamefont {Turek},
  \citenamefont {Drchal}, \citenamefont {Kudrnovsk\'y}, \citenamefont
  {\v{S}ob},\ and\ \citenamefont {Weinberger}}]{r_1997_tdk}%
  \BibitemOpen
  \bibfield  {author} {\bibinfo {author} {\bibfnamefont {I.}~\bibnamefont
  {Turek}}, \bibinfo {author} {\bibfnamefont {V.}~\bibnamefont {Drchal}},
  \bibinfo {author} {\bibfnamefont {J.}~\bibnamefont {Kudrnovsk\'y}}, \bibinfo
  {author} {\bibfnamefont {M.}~\bibnamefont {\v{S}ob}}, \ and\ \bibinfo
  {author} {\bibfnamefont {P.}~\bibnamefont {Weinberger}},\ }\href@noop {}
  {\emph {\bibinfo {title} {Electronic Structure of Disordered Alloys, Surfaces
  and Interfaces}}}\ (\bibinfo  {publisher} {Kluwer, Boston},\ \bibinfo {year}
  {1997})\BibitemShut {NoStop}%
\bibitem [{\citenamefont {Weinberger}(1990)}]{r_1990_pw}%
  \BibitemOpen
  \bibfield  {author} {\bibinfo {author} {\bibfnamefont {P.}~\bibnamefont
  {Weinberger}},\ }\href@noop {} {\emph {\bibinfo {title} {Electron Scattering
  Theory for Ordered and Disordered Matter}}}\ (\bibinfo  {publisher}
  {Clarendon Press, Oxford},\ \bibinfo {year} {1990})\BibitemShut {NoStop}%
\bibitem [{\citenamefont {Zabloudil}\ \emph {et~al.}(2005)\citenamefont
  {Zabloudil}, \citenamefont {Hammerling}, \citenamefont {Szunyogh},\ and\
  \citenamefont {Weinberger}}]{r_2005_zhs}%
  \BibitemOpen
  \bibfield  {author} {\bibinfo {author} {\bibfnamefont {J.}~\bibnamefont
  {Zabloudil}}, \bibinfo {author} {\bibfnamefont {R.}~\bibnamefont
  {Hammerling}}, \bibinfo {author} {\bibfnamefont {L.}~\bibnamefont
  {Szunyogh}}, \ and\ \bibinfo {author} {\bibfnamefont {P.}~\bibnamefont
  {Weinberger}},\ }\href@noop {} {\emph {\bibinfo {title} {Electron Scattering
  in Solid Matter}}}\ (\bibinfo  {publisher} {Springer, Berlin},\ \bibinfo
  {year} {2005})\BibitemShut {NoStop}%
\bibitem [{\citenamefont {Turek}\ \emph {et~al.}(2002)\citenamefont {Turek},
  \citenamefont {Kudrnovsk\'y}, \citenamefont {Drchal}, \citenamefont
  {Szunyogh},\ and\ \citenamefont {Weinberger}}]{r_2002_tkd}%
  \BibitemOpen
  \bibfield  {author} {\bibinfo {author} {\bibfnamefont {I.}~\bibnamefont
  {Turek}}, \bibinfo {author} {\bibfnamefont {J.}~\bibnamefont {Kudrnovsk\'y}},
  \bibinfo {author} {\bibfnamefont {V.}~\bibnamefont {Drchal}}, \bibinfo
  {author} {\bibfnamefont {L.}~\bibnamefont {Szunyogh}}, \ and\ \bibinfo
  {author} {\bibfnamefont {P.}~\bibnamefont {Weinberger}},\ }\href@noop {}
  {\bibfield  {journal} {\bibinfo  {journal} {Phys. Rev. B}\ }\textbf {\bibinfo
  {volume} {65}},\ \bibinfo {pages} {125101} (\bibinfo {year}
  {2002})}\BibitemShut {NoStop}%
\bibitem [{\citenamefont {Turek}\ \emph {et~al.}(2012)\citenamefont {Turek},
  \citenamefont {Kudrnovsk\'y},\ and\ \citenamefont {Drchal}}]{r_2012_tkd}%
  \BibitemOpen
  \bibfield  {author} {\bibinfo {author} {\bibfnamefont {I.}~\bibnamefont
  {Turek}}, \bibinfo {author} {\bibfnamefont {J.}~\bibnamefont {Kudrnovsk\'y}},
  \ and\ \bibinfo {author} {\bibfnamefont {V.}~\bibnamefont {Drchal}},\
  }\href@noop {} {\bibfield  {journal} {\bibinfo  {journal} {Phys. Rev. B}\
  }\textbf {\bibinfo {volume} {86}},\ \bibinfo {pages} {014405} (\bibinfo
  {year} {2012})}\BibitemShut {NoStop}%
\bibitem [{\citenamefont {Turek}\ \emph {et~al.}(2014)\citenamefont {Turek},
  \citenamefont {Kudrnovsk\'y},\ and\ \citenamefont {Drchal}}]{r_2014_tkd}%
  \BibitemOpen
  \bibfield  {author} {\bibinfo {author} {\bibfnamefont {I.}~\bibnamefont
  {Turek}}, \bibinfo {author} {\bibfnamefont {J.}~\bibnamefont {Kudrnovsk\'y}},
  \ and\ \bibinfo {author} {\bibfnamefont {V.}~\bibnamefont {Drchal}},\
  }\href@noop {} {\bibfield  {journal} {\bibinfo  {journal} {Phys. Rev. B}\
  }\textbf {\bibinfo {volume} {89}},\ \bibinfo {pages} {064405} (\bibinfo
  {year} {2014})}\BibitemShut {NoStop}%
\bibitem [{\citenamefont {Kudrnovsk\'y}\ and\ \citenamefont
  {Drchal}(1990)}]{r_1990_kd}%
  \BibitemOpen
  \bibfield  {author} {\bibinfo {author} {\bibfnamefont {J.}~\bibnamefont
  {Kudrnovsk\'y}}\ and\ \bibinfo {author} {\bibfnamefont {V.}~\bibnamefont
  {Drchal}},\ }\href@noop {} {\bibfield  {journal} {\bibinfo  {journal} {Phys.
  Rev. B}\ }\textbf {\bibinfo {volume} {41}},\ \bibinfo {pages} {7515}
  (\bibinfo {year} {1990})}\BibitemShut {NoStop}%
\bibitem [{\citenamefont {Carva}\ \emph {et~al.}(2006)\citenamefont {Carva},
  \citenamefont {Turek}, \citenamefont {Kudrnovsk\'y},\ and\ \citenamefont
  {Bengone}}]{r_2006_ctk}%
  \BibitemOpen
  \bibfield  {author} {\bibinfo {author} {\bibfnamefont {K.}~\bibnamefont
  {Carva}}, \bibinfo {author} {\bibfnamefont {I.}~\bibnamefont {Turek}},
  \bibinfo {author} {\bibfnamefont {J.}~\bibnamefont {Kudrnovsk\'y}}, \ and\
  \bibinfo {author} {\bibfnamefont {O.}~\bibnamefont {Bengone}},\ }\href@noop
  {} {\bibfield  {journal} {\bibinfo  {journal} {Phys. Rev. B}\ }\textbf
  {\bibinfo {volume} {73}},\ \bibinfo {pages} {144421} (\bibinfo {year}
  {2006})}\BibitemShut {NoStop}%
\bibitem [{\citenamefont {Andersen}\ and\ \citenamefont
  {Jepsen}(1984)}]{r_1984_aj}%
  \BibitemOpen
  \bibfield  {author} {\bibinfo {author} {\bibfnamefont {O.~K.}\ \bibnamefont
  {Andersen}}\ and\ \bibinfo {author} {\bibfnamefont {O.}~\bibnamefont
  {Jepsen}},\ }\href@noop {} {\bibfield  {journal} {\bibinfo  {journal} {Phys.
  Rev. Lett.}\ }\textbf {\bibinfo {volume} {53}},\ \bibinfo {pages} {2571}
  (\bibinfo {year} {1984})}\BibitemShut {NoStop}%
\bibitem [{\citenamefont {Andersen}\ \emph {et~al.}(1986)\citenamefont
  {Andersen}, \citenamefont {Pawlowska},\ and\ \citenamefont
  {Jepsen}}]{r_1986_apj}%
  \BibitemOpen
  \bibfield  {author} {\bibinfo {author} {\bibfnamefont {O.~K.}\ \bibnamefont
  {Andersen}}, \bibinfo {author} {\bibfnamefont {Z.}~\bibnamefont {Pawlowska}},
  \ and\ \bibinfo {author} {\bibfnamefont {O.}~\bibnamefont {Jepsen}},\
  }\href@noop {} {\bibfield  {journal} {\bibinfo  {journal} {Phys. Rev. B}\
  }\textbf {\bibinfo {volume} {34}},\ \bibinfo {pages} {5253} (\bibinfo {year}
  {1986})}\BibitemShut {NoStop}%
\bibitem [{\citenamefont {Hatherly}\ \emph {et~al.}(1964)\citenamefont
  {Hatherly}, \citenamefont {Hirakawa}, \citenamefont {Lowde}, \citenamefont
  {Mallett}, \citenamefont {Stringfellow},\ and\ \citenamefont
  {Torrie}}]{r_1964_hhl}%
  \BibitemOpen
  \bibfield  {author} {\bibinfo {author} {\bibfnamefont {M.}~\bibnamefont
  {Hatherly}}, \bibinfo {author} {\bibfnamefont {K.}~\bibnamefont {Hirakawa}},
  \bibinfo {author} {\bibfnamefont {R.~D.}\ \bibnamefont {Lowde}}, \bibinfo
  {author} {\bibfnamefont {J.~F.}\ \bibnamefont {Mallett}}, \bibinfo {author}
  {\bibfnamefont {M.~W.}\ \bibnamefont {Stringfellow}}, \ and\ \bibinfo
  {author} {\bibfnamefont {B.~H.}\ \bibnamefont {Torrie}},\ }\href@noop {}
  {\bibfield  {journal} {\bibinfo  {journal} {Proc. Phys. Soc. London}\
  }\textbf {\bibinfo {volume} {84}},\ \bibinfo {pages} {55} (\bibinfo {year}
  {1964})}\BibitemShut {NoStop}%
\bibitem [{\citenamefont {Pauthenet}(1982)}]{r_1982_rp_b}%
  \BibitemOpen
  \bibfield  {author} {\bibinfo {author} {\bibfnamefont {R.}~\bibnamefont
  {Pauthenet}},\ }\href@noop {} {\bibfield  {journal} {\bibinfo  {journal} {J.
  Appl. Phys.}\ }\textbf {\bibinfo {volume} {53}},\ \bibinfo {pages} {8187}
  (\bibinfo {year} {1982})}\BibitemShut {NoStop}%
\bibitem [{\citenamefont {Nakai}(1983)}]{r_1983_in}%
  \BibitemOpen
  \bibfield  {author} {\bibinfo {author} {\bibfnamefont {I.}~\bibnamefont
  {Nakai}},\ }\href@noop {} {\bibfield  {journal} {\bibinfo  {journal} {J.
  Phys. Soc. Jpn.}\ }\textbf {\bibinfo {volume} {52}},\ \bibinfo {pages} {1781}
  (\bibinfo {year} {1983})}\BibitemShut {NoStop}%
\bibitem [{\citenamefont {Eastman}\ \emph {et~al.}(1980)\citenamefont
  {Eastman}, \citenamefont {Himpsel},\ and\ \citenamefont
  {Knapp}}]{r_1980_ehk}%
  \BibitemOpen
  \bibfield  {author} {\bibinfo {author} {\bibfnamefont {D.~E.}\ \bibnamefont
  {Eastman}}, \bibinfo {author} {\bibfnamefont {F.~J.}\ \bibnamefont
  {Himpsel}}, \ and\ \bibinfo {author} {\bibfnamefont {J.~A.}\ \bibnamefont
  {Knapp}},\ }\href@noop {} {\bibfield  {journal} {\bibinfo  {journal} {Phys.
  Rev. Lett.}\ }\textbf {\bibinfo {volume} {44}},\ \bibinfo {pages} {95}
  (\bibinfo {year} {1980})}\BibitemShut {NoStop}%
\bibitem [{\citenamefont {Katsnelson}\ and\ \citenamefont
  {Lichtenstein}(2000)}]{r_2000_kl}%
  \BibitemOpen
  \bibfield  {author} {\bibinfo {author} {\bibfnamefont {M.~I.}\ \bibnamefont
  {Katsnelson}}\ and\ \bibinfo {author} {\bibfnamefont {A.~I.}\ \bibnamefont
  {Lichtenstein}},\ }\href@noop {} {\bibfield  {journal} {\bibinfo  {journal}
  {Phys. Rev. B}\ }\textbf {\bibinfo {volume} {61}},\ \bibinfo {pages} {8906}
  (\bibinfo {year} {2000})}\BibitemShut {NoStop}%
\bibitem [{\citenamefont {Lowde}\ \emph {et~al.}(1965)\citenamefont {Lowde},
  \citenamefont {Shimizu}, \citenamefont {Stringfellow},\ and\ \citenamefont
  {Torrie}}]{r_1965_lss}%
  \BibitemOpen
  \bibfield  {author} {\bibinfo {author} {\bibfnamefont {R.~D.}\ \bibnamefont
  {Lowde}}, \bibinfo {author} {\bibfnamefont {M.}~\bibnamefont {Shimizu}},
  \bibinfo {author} {\bibfnamefont {M.~W.}\ \bibnamefont {Stringfellow}}, \
  and\ \bibinfo {author} {\bibfnamefont {B.~H.}\ \bibnamefont {Torrie}},\
  }\href@noop {} {\bibfield  {journal} {\bibinfo  {journal} {Phys. Rev. Lett.}\
  }\textbf {\bibinfo {volume} {14}},\ \bibinfo {pages} {698} (\bibinfo {year}
  {1965})}\BibitemShut {NoStop}%
\bibitem [{\citenamefont {Antonini}\ and\ \citenamefont
  {Stringfellow}(1966)}]{r_1966_as}%
  \BibitemOpen
  \bibfield  {author} {\bibinfo {author} {\bibfnamefont {B.}~\bibnamefont
  {Antonini}}\ and\ \bibinfo {author} {\bibfnamefont {M.~W.}\ \bibnamefont
  {Stringfellow}},\ }\href@noop {} {\bibfield  {journal} {\bibinfo  {journal}
  {Proc. Phys. Soc. London}\ }\textbf {\bibinfo {volume} {89}},\ \bibinfo
  {pages} {419} (\bibinfo {year} {1966})}\BibitemShut {NoStop}%
\bibitem [{\citenamefont {Yelsukov}\ \emph {et~al.}(1992)\citenamefont
  {Yelsukov}, \citenamefont {Voronina},\ and\ \citenamefont
  {Barinov}}]{r_1992_yvb}%
  \BibitemOpen
  \bibfield  {author} {\bibinfo {author} {\bibfnamefont {E.~P.}\ \bibnamefont
  {Yelsukov}}, \bibinfo {author} {\bibfnamefont {E.~V.}\ \bibnamefont
  {Voronina}}, \ and\ \bibinfo {author} {\bibfnamefont {V.~A.}\ \bibnamefont
  {Barinov}},\ }\href@noop {} {\bibfield  {journal} {\bibinfo  {journal} {J.
  Magn. Magn. Mater.}\ }\textbf {\bibinfo {volume} {115}},\ \bibinfo {pages}
  {271} (\bibinfo {year} {1992})}\BibitemShut {NoStop}%
\bibitem [{\citenamefont {Kleykamp}\ and\ \citenamefont
  {Glasbrenner}(1997)}]{r_1997_kg}%
  \BibitemOpen
  \bibfield  {author} {\bibinfo {author} {\bibfnamefont {H.}~\bibnamefont
  {Kleykamp}}\ and\ \bibinfo {author} {\bibfnamefont {H.}~\bibnamefont
  {Glasbrenner}},\ }\href@noop {} {\bibfield  {journal} {\bibinfo  {journal}
  {Z. Metallkd.}\ }\textbf {\bibinfo {volume} {88}},\ \bibinfo {pages} {230}
  (\bibinfo {year} {1997})}\BibitemShut {NoStop}%
\bibitem [{\citenamefont {Massalski}(1986)}]{r_1986_tbm}%
  \BibitemOpen
  \bibinfo {editor} {\bibfnamefont {T.~B.}\ \bibnamefont {Massalski}},\ ed.,\
  \href@noop {} {\emph {\bibinfo {title} {Binary Alloy Phase Diagrams}}}\
  (\bibinfo  {publisher} {American Society for Metals, Metals Park, Ohio},\
  \bibinfo {year} {1986})\BibitemShut {NoStop}%
\bibitem [{\citenamefont {Frikkee}(1978)}]{r_1978_ef}%
  \BibitemOpen
  \bibfield  {author} {\bibinfo {author} {\bibfnamefont {E.}~\bibnamefont
  {Frikkee}},\ }\href@noop {} {\bibfield  {journal} {\bibinfo  {journal} {J.
  Phys. F: Met. Phys.}\ }\textbf {\bibinfo {volume} {8}},\ \bibinfo {pages}
  {L141} (\bibinfo {year} {1978})}\BibitemShut {NoStop}%
\bibitem [{\citenamefont {Staunton}\ \emph {et~al.}(1997)\citenamefont
  {Staunton}, \citenamefont {Ling},\ and\ \citenamefont
  {Johnson}}]{r_1997_slj}%
  \BibitemOpen
  \bibfield  {author} {\bibinfo {author} {\bibfnamefont {J.~B.}\ \bibnamefont
  {Staunton}}, \bibinfo {author} {\bibfnamefont {M.~F.}\ \bibnamefont {Ling}},
  \ and\ \bibinfo {author} {\bibfnamefont {D.~D.}\ \bibnamefont {Johnson}},\
  }\href@noop {} {\bibfield  {journal} {\bibinfo  {journal} {J. Phys.: Condens.
  Matter}\ }\textbf {\bibinfo {volume} {9}},\ \bibinfo {pages} {1281} (\bibinfo
  {year} {1997})}\BibitemShut {NoStop}%
\bibitem [{\citenamefont {Staunton}\ \emph {et~al.}(1998)\citenamefont
  {Staunton}, \citenamefont {Razee}, \citenamefont {Ling}, \citenamefont
  {Johnson},\ and\ \citenamefont {Pinski}}]{r_1998_srl}%
  \BibitemOpen
  \bibfield  {author} {\bibinfo {author} {\bibfnamefont {J.~B.}\ \bibnamefont
  {Staunton}}, \bibinfo {author} {\bibfnamefont {S.~S.~A.}\ \bibnamefont
  {Razee}}, \bibinfo {author} {\bibfnamefont {M.~F.}\ \bibnamefont {Ling}},
  \bibinfo {author} {\bibfnamefont {D.~D.}\ \bibnamefont {Johnson}}, \ and\
  \bibinfo {author} {\bibfnamefont {F.~J.}\ \bibnamefont {Pinski}},\
  }\href@noop {} {\bibfield  {journal} {\bibinfo  {journal} {J. Phys. D: Appl.
  Phys.}\ }\textbf {\bibinfo {volume} {31}},\ \bibinfo {pages} {2355} (\bibinfo
  {year} {1998})}\BibitemShut {NoStop}%
\bibitem [{\citenamefont {Swihart}\ \emph {et~al.}(1986)\citenamefont
  {Swihart}, \citenamefont {Butler}, \citenamefont {Stocks}, \citenamefont
  {Nicholson},\ and\ \citenamefont {Ward}}]{r_1986_sbs}%
  \BibitemOpen
  \bibfield  {author} {\bibinfo {author} {\bibfnamefont {J.~C.}\ \bibnamefont
  {Swihart}}, \bibinfo {author} {\bibfnamefont {W.~H.}\ \bibnamefont {Butler}},
  \bibinfo {author} {\bibfnamefont {G.~M.}\ \bibnamefont {Stocks}}, \bibinfo
  {author} {\bibfnamefont {D.~M.}\ \bibnamefont {Nicholson}}, \ and\ \bibinfo
  {author} {\bibfnamefont {R.~C.}\ \bibnamefont {Ward}},\ }\href@noop {}
  {\bibfield  {journal} {\bibinfo  {journal} {Phys. Rev. Lett.}\ }\textbf
  {\bibinfo {volume} {57}},\ \bibinfo {pages} {1181} (\bibinfo {year}
  {1986})}\BibitemShut {NoStop}%
\bibitem [{\citenamefont {Antropov}\ and\ \citenamefont
  {Liechtenstein}(1992)}]{r_1992_al}%
  \BibitemOpen
  \bibfield  {author} {\bibinfo {author} {\bibfnamefont {V.~P.}\ \bibnamefont
  {Antropov}}\ and\ \bibinfo {author} {\bibfnamefont {A.~I.}\ \bibnamefont
  {Liechtenstein}},\ }\href@noop {} {\bibfield  {journal} {\bibinfo  {journal}
  {Mater. Res. Soc. Symp. Proc.}\ }\textbf {\bibinfo {volume} {253}},\ \bibinfo
  {pages} {325} (\bibinfo {year} {1992})}\BibitemShut {NoStop}%
\bibitem [{\citenamefont {Udvardi}\ \emph {et~al.}(2003)\citenamefont
  {Udvardi}, \citenamefont {Szunyogh}, \citenamefont {Palot\'as},\ and\
  \citenamefont {Weinberger}}]{r_2003_usp}%
  \BibitemOpen
  \bibfield  {author} {\bibinfo {author} {\bibfnamefont {L.}~\bibnamefont
  {Udvardi}}, \bibinfo {author} {\bibfnamefont {L.}~\bibnamefont {Szunyogh}},
  \bibinfo {author} {\bibfnamefont {K.}~\bibnamefont {Palot\'as}}, \ and\
  \bibinfo {author} {\bibfnamefont {P.}~\bibnamefont {Weinberger}},\
  }\href@noop {} {\bibfield  {journal} {\bibinfo  {journal} {Phys. Rev. B}\
  }\textbf {\bibinfo {volume} {68}},\ \bibinfo {pages} {104436} (\bibinfo
  {year} {2003})}\BibitemShut {NoStop}%
\end{thebibliography}

\providecommand{\noopsort}[1]{}\providecommand{\singleletter}[1]{#1}%

\end{document}